\documentstyle[aps,preprint]{revtex}
\begin{document}
 \tighten
\input epsf
\draft
\preprint{hep-th/0104073}
\date{April 24, 2001}
\title{\Large\bf Pyrotechnic Universe}
\author{Renata Kallosh,$^1$  Lev Kofman$^2$  and  Andrei Linde$^1$} \address{$^1$ Department of Physics, Stanford University, Stanford, CA 94305-4060,
USA} \address{$^2$ CITA, University of Toronto, 60 St. George Street,
Toronto, ON M5S 3H8, Canada} \maketitle
\begin{abstract}
One of the central points of the ekpyrotic cosmological scenario based on
Ho\u{r}ava-Witten theory is that we live on a negative tension brane.
However, the  tension of the visible brane is positive in the usual HW phenomenology
with stronger coupling on the hidden brane, both for standard and 
non-standard embedding. To make ekpyrotic
scenario realistic one must solve  the problem of the negative
cosmological constant on the visible brane and fine-tune the  bulk brane
potential with an accuracy of $10^{-50}$. In terms of a canonically
normalized scalar field $\phi$ describing the position of the brane, this
potential must take a very unusual form ${\cal V}(\phi)= -10^{-22} M_p^4
\exp \left(-{5000\, \phi\over M_p}\right)$. We describe the problems which
appear when one attempts to obtain this potential in string theory.  The
mechanism for the generation of density perturbations in this scenario is
not brane-specific; it is a particular limiting case of the mechanism of
tachyonic preheating.  Unlike inflation, this mechanism exponentially
amplifies not only quantum fluctuations, but also initial inhomogeneities.
As a result, to solve the homogeneity problem in this scenario, one would
need the branes to be parallel to each other with an accuracy better than
$10^{-60}$ on a scale $10^{30}$ times greater than the distance between
the branes. Thus, at present, inflation remains the only robust mechanism
that produces density perturbations with a flat spectrum and
simultaneously solves all major cosmological problems.
\end{abstract}

\section{Introduction}
After 15 years of development of string theory and M-theory we are still
faced with the challenging problem of constructing a consistent and
realistic stringy cosmology.  An interesting step in this direction has
recently been made by Khoury,  Ovrut,  Steinhardt and  Turok, who
suggested a three-brane cosmological model based on the Ho\u{r}ava-Witten 
theory, and argued that it may resolve
all major cosmological problems without any use of inflation
\cite{KOST}. Their model was called the ekpyrotic universe, from the
Greek-derived word {\it ekpyrosis}.

The basic idea of this scenario is that initially the universe was in a
nearly BPS state consisting of two parallel branes, and that we live on
the brane with negative tension. Then the brane with positive tension
splits into two positive tension branes, one of which (bulk brane) starts
moving towards our brane. The big bang corresponds to the moment when the
bulk brane hits our brane; the collision makes the universe hot. It was
argued that the flatness of the branes in the nearly BPS state solves the
homogeneity and flatness problems, whereas quantum fluctuations of the
bulk brane result in the large-scale density perturbations on our brane
when these branes collide. In this paper we re-examine some of the basic
premises of this model.  We will try to verify whether it follows from
string theory and whether it indeed  can solve all major cosmological
problems without the help of inflation.  In this sense, our paper will be
devoted to an epicrisis of ekpyrosis.\footnote{Epicrisis is a Greek word
for critical evaluation.}

One of the central points of the ekpyrotic scenario is that we live on a negative tension
brane, and the warp factor (the volume of the Calabi-Yau space) decreases towards the visible brane.
In the original version of Ref. \cite{KOST} one can read: {\it As we   
will see in Section VB,   it will be   necessary  for the visible brane to   
be in the small-volume region of space-time.} The authors repeatedly emphasized that this condition is very important for their scenario and argued that it results in a distinguishing feature
of their model: a blue spectrum of density perturbations.\footnote{They
also noticed that gravitational waves  in the ekpyrotic scenario will have a
strongly blue spectrum; but, since their level is going to be extremely
small, the shape of the spectrum of the gravitational waves will be nearly
impossible to determine.}

However, as we will explain in Section \ref{super},  the standard HW phenomenology \cite{Witten,HWph,lukas} (both for standard and non-standard embedding) is based  on the assumption that the tension of the visible brane is positive, and   the warp factor increases towards the visible brane.  There were two main reasons for such an assumption. First of all, in practically all known  versions of the HW phenomenology,  with few exceptions,  a smaller group of symmetry (such as $E_6$) lives on the positive tension brane and provides the basis for GUTs, whereas the symmetry $E_8$ on the negative tension brane may remain unbroken. It is very difficult   to find  models where $E_6$ or $SU(5)$ live on the negative tension brane \cite{Benakli:1999sy,Donagi:2001fs}.  

There is another reason why the tension of the visible brane is positive in the standard HW phenomenology \cite{Witten,HWph,lukas}:  The square of the gauge coupling constant is inversely proportional to the Calabi-Yau volume \cite{Witten}. On the negative tension brane this volume is greater than on the positive tension one, see e.g. \cite{KOST}. In the standard HW phenomenology it is usually assumed that we live on the positive tension brane with small gauge coupling, ${g^2_{GUT}\over 4\pi} \sim 0.04$. On the hidden brane with negative tension the gauge coupling constant becomes large, ${g^2_{hidden}\over 4\pi} = O(1)$, which makes the gaugino condensation possible \cite{Witten,HWph,lukas}. It is not  impossible to have a consistent phenomenology with the small gauge coupling on the hidden brane, but this is an unconventional and not well explored possibility \cite{Benakli:1999sy}.

Thus, we believe that the ekpyrotic scenario is at odds with the standard HW phenomenology as defined in \cite{Witten,HWph,lukas}. The relevant issue is not the standard versus non-standard embedding,  but Ho\u{r}ava-Witten phenomenology \cite{Witten,HWph,lukas} versus   
Benakli-Lalak-Pokorski-Thomas \cite{Benakli:1999sy} phenomenology.
\footnote{Note that one should clearly distinguish between the non-standard phenomenology of \cite{KOST} and the non-standard embedding that is required to describe the bulk brane.}  
As explained in Section VB of \cite{KOST}, the reason to assume that the CY volume should decrease towards the visible brane was rooted in the idea that this is required for generation of density perturbations in the ekpyrotic scenario. However, as we will show in Section \ref{perturb}, this requirement is not necessary.

To improve  the ekpyrotic scenario one would need to change the sign of the brane tension. In this case the warp factor decreases towards the
hidden brane. This changes the shape of the spectrum of density
perturbations from blue to red. However, one cannot simply flip the sign
of the tension in the model leaving all other parameters intact because it
would introduce a singularity between the branes.  One needs to change
other parameters of the model as well. We will call the improved scenario
{\it pyrotechnic} to emphasize its relation to the ekpyrotic scenario, but
also to indicate that it remains vulnerable to other problems to be
discussed below. In this scenario, unlike in the ekpyrotic scenario, we will not make any attempts to avoid inflation.

The critical  assumption of the ekpyrotic scenario discussed in Section \ref{pot} is that the bulk brane interacts with the visible brane with the negative potential
\begin{equation}
  V(Y)= -v e^{- |\alpha|m Y} \ .
\label{pot1}
\end{equation}
However, in  cases where this potential has been
explicitly calculated, it was shown to be positive \cite{Moore:2000fs}.
Moreover, in general the potential contains two terms, $e^{-|\alpha| m Y}$ and 
$e^{- |\alpha| m(R-Y)}$, to include bulk brane interactions with hidden and visible branes.
Originally, potentials of such type were supposed to describe interactions of the visible brane or the bulk brane with the hidden brane with unbroken $E_8$ \cite{LOP}. In such a case the potential would be $V(Y) \sim -v  e^{- |\alpha|m (R-Y)}$, where $R-Y$ is the distance between the bulk brane and the hidden brane. In the ekpyrotic scenario all terms like that should be forbidden, which may be difficult to achieve unless one assumes that originally our brane was the end-of-the-world $E_8$ brane, and it became the physical brane after the brane collision. No theoretical description of such a process is presently available; all previous attempts to do so assumed that $E_8$ is already broken and colliding branes have comparable tensions, which is not the case in \cite{KOST}.
 
The peculiar nature of the brane potential $V(Y)
\sim -v  e^{-|\alpha | m Y}$ becomes manifest if one writes it in terms of a canonically normalized scalar field $\phi$ describing the position of the brane: ${\cal V}(\phi)= -10^{-22} M_p^4 \exp \left(-{5000\, \phi\over M_p}\right)$ for the specific choice of the 
parameters $v$, $\alpha$, $m$  requested by \cite{KOST}. While potentials  $\sim \exp {c\phi\over M_p}$ with $c = O(1)$ often appear in string theory, such terms as $\exp \left(-{5000\, \phi\over M_p}\right)$ are rather unprecedented.

At the first glance, the theory of the generation of density perturbations
in the ekpyrotic scenario seems very complex and brane-specific, as
indicated by the statement of \cite{KOST} that this mechanism requires the
warp factor to decrease towards the visible brane. However, in Section
\ref{perturb} we show that the theory of the generation of density
perturbations used in \cite{KOST} is in fact a limiting case of the theory
of tachyonic preheating recently developed in \cite{tach}.  The theory of
this effect is very simple. It works for branes in 5d as well as  for the
usual scalar field in 4d due to the exponential growth of long wavelength
fluctuations  in theories with concave effective potentials ($V'' < 0$).
The spectrum of perturbations may be flat, but it may also be red or blue,
depending on the choice of the potential. Power-law potentials typically
are unacceptable. One should make the very special choice of a nearly
exponential potential  to produce a cosmologically acceptable spectrum. To
make this mechanism realistic, one must solve  the problem of the negative
cosmological constant on the visible brane and fine-tune the  value of the
bulk brane potential with an accuracy of $10^{-50}$. If one does not
perform this fine-tuning, the standard inflationary mechanism for the
generation of density perturbations turns on, and the model becomes very
similar to the model of brane inflation proposed by Dvali and Tye
\cite{DvaliTye}.

If one resolves all of these problems, and the tachyonic mechanism for
the generation of density perturbations begins to work, then  we will have a new problem, which may be much more serious than the previous ones: tachyonic instability
exponentially amplifies not only quantum fluctuations but also initial
inhomogeneities. As a result, to solve the homogeneity problem in this
scenario one would need the branes to be parallel to each other with  an
accuracy  better than $10^{-60}$ on a scale $10^{30}$ times greater than
the distance between the branes, see Section \ref{hom}. Since the initial
state is not really a true BPS state but rather some unstable and evolving
configuration, we do not see any reason why our universe must be so
incredibly homogeneous from the very beginning. 

Thus we believe that at present inflation remains the only robust
mechanism that produces density perturbations with a flat spectrum and
simultaneously solves all major cosmological problems.

\section{General setup for ekpyrotic universe}\label{setup}

The ekpyrotic scenario consists of many parts  related to the M-theory and cosmology. M-theory issues, including the sign of the tensions of the
visible and hidden brane and the BPS nature of the 3 brane solution, will
be discussed in Sec. 3. For a cosmologist, the end result of the story
from M-theory as presented in \cite{KOST} is the following:

There is a  static  three brane solution for the space-time metric and the
dilaton $e^{\phi}$ (volume of the Calabi-Yau space) given by
\begin{eqnarray}
\nonumber & & ds^2=D(y)(-N^2d\tau^2+A^2d\vec{x}^2)+B^2 D^4(y)dy^2 \ ,
\nonumber\\ & & e^{\phi}=B D^3(y) \ , \nonumber\\ & & D(y)=\alpha y +C
\;\;\;\;\;\;\;\;\;\;\;\;\;\;\;\;\;\;\;\;\;\;\;\;\;\;{\rm
for}\;\;y<Y \\
& & \;\;\;\;\;\;\;\;\;=(\alpha-\beta)y+C+\beta Y \;\;\;\;\;\;\;{\rm
for}\;\;y>Y, \label{eq:static}
\end{eqnarray}
where $A,B,C,N$  are constants and $C>0$. The boundary branes are located
at $y=0$ and $y=R$, and the bulk brane is located at $y=Y$, where $0\leq
Y\leq R$.  The tension of the visible brane at $y=0$ is $-\alpha$ and is
{\it negative}.  The tension of the bulk brane $\beta$ is positive and the
tension of the hidden brane at $y=R$ is  positive and equals
$\alpha-\beta$. One assumes that $\beta \ll \alpha$, so the bulk brane is
relatively light. The visible brane at $y=0$ lies in the region of smaller
volume  while $y=R$ lies in the region of larger volume. Indeed, $D(0)= C$
and $D(R)= C + \alpha R$ and  $\alpha$ is positive, so $D(0) < D(R)$. This
property is considered one of the most important features of the scenario.

The light bulk brane may either appear spontaneously from the hidden brane
or it may also exist from the very beginning, i.e. one starts with two
boundary branes and one bulk brane. The three brane configuration is
assumed to be in a nearly BPS state. It is argued that the universe must
be homogeneous because the BPS brane configuration is homogeneous. The
bulk brane has a kinetic term and a potential, which  for a ``successful
example'' is chosen to be $V(Y)= -v e^{-m\alpha Y}$. Additionally  it is
assumed that at small $Y$ the potential suddenly becomes zero due to some
nonperturbative effects.

 Due to the slight contraction of the scale factor on the
bulk  brane, the bulk  brane carries some residual kinetic energy
immediately before the collision with the visible brane. After the
collision, this residual kinetic energy transforms into radiation which
will be deposited in the three dimensional space of the visible  brane.
The visible brane, now filled with hot radiation, somehow begins to expand
as a flat FRW universe. However, the temperature is not high enough to
trigger phase transitions in GUTs and produce primordial monopoles.
Quantum fluctuations of the position of the bulk brane generated during
its  motion from $Y= R$ to $Y=0$  will result in  density fluctuations
with a nearly flat spectrum.  The spectrum will have a slightly blue tilt
for the exponential potential $V(Y)$.

The set of parameters used in \cite{KOST} is $\alpha=250M_5$,
$\beta=0.1M_5$, $B=10^{-3}$, $C=100$, $R=M_5^{-1}$, $v\sim 10^{-10}$,
$m=1$, and $M_5=10^{-2}M_{pl}$. A sketch of the model is given in Fig. 1.

The total setup is rather complicated, but the final picture allows for a
dramatic simplification. Indeed, let us look first at the behavior of the
factor $D(Y)$, which determines the metric in (\ref{eq:static}), during
the whole process of motion of the bulk brane towards the visible brane.
The motion begins at $Y = R =  M_5^{-1}$ and ends at $Y=0$. During this
process $D(Y)$ changes from $350$ to $100$, which is not that much. A more
complicated analysis performed in  \cite{KOST} shows that the scale
factors on all branes also do not change much. This suggests that the
expansion of the universe and other complicated gravitational effects
cannot be of any relevance to the possibility to solve major cosmological
problems and to the basic mechanism of the generation of density
perturbations. On the other hand, the authors of Ref. \cite{KOST}
emphasized that the decrease of $D(Y)$ towards the brane is crucially
important and used the notion of the effective scale factor $a_{\rm
{eff}}$ to explain the mechanism of production of density perturbations.

In this paper we will attempt to analyse this situation, starting from the
M-theory part, and ending with a discussion of the cosmological density
perturbations and homogeneity problem.

\begin{figure}[b]
\centering\leavevmode\epsfysize=8cm \epsfbox{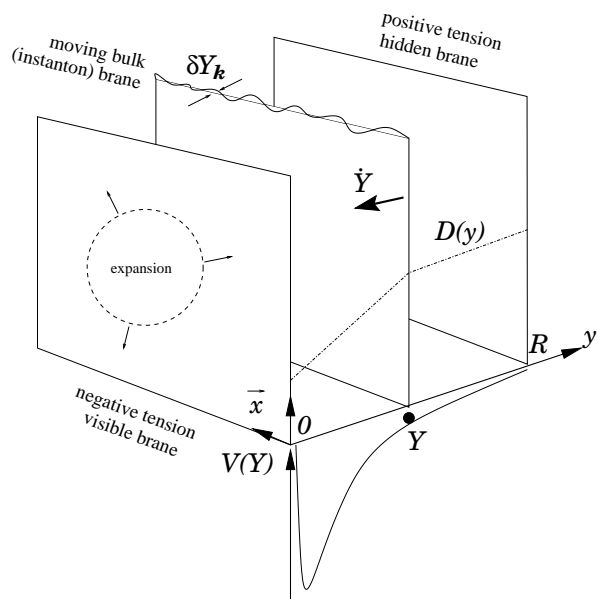}
\caption[fig1]{\label{fig1} Sketch of the  ekpyrotic  scenario. We live on
a brane with {  negative} energy density. The Big Bang occurs when the
bulk brane hits our brane. The bulk brane has potential energy $V(Y)$
which is postulated to have a very specific form: it is negative
everywhere except on our brane, and its absolute value decreases
exponentially at large $Y$. An important feature of this scenario is that
the volume of space, controlled by the metric $D(Y)$, decreases near our
brane, which makes the spectrum of perturbations blue.}
\end{figure}

\section{Supersymmetry and  hidden-visible branes}\label{super}

\subsection{The action and the static solution}
Here we  consider the specific construction of \cite{KOST} which starts
with the solution  of two boundary branes with almost opposite tensions
and a bulk brane in between.  We will explain  the reason why the status
of unbroken supersymmetry (BPS property) of this solution is problematic.

An effective five-dimensional action of heterotic M-theory is   given in
\cite{KOST}
the form:

\begin{eqnarray}
\nonumber & & S=\frac{M_5^3}{2}\int_{{\cal M}_5}
d^5x\sqrt{-g}\left(R-\frac{1}{2}(\partial\phi)^2-\frac{e^{2\phi}{\cal
F}^2}{5!}\right) \\
& &-3\sum_{i=1}^3\alpha_iM_5^3\int_{{\cal M}_4^{(i)}}
d^4\xi_{(i)}\Bigl(\sqrt{-h_{(i)}}e^{-\phi}
 -\frac{\epsilon^{\mu\nu\kappa\lambda}}{4!}{\cal
A}_{\gamma\delta\epsilon\zeta}
\partial_{\mu}X^{\gamma}_{(i)}\partial_{\nu}X^{\delta}_{(i)}
\partial_{\kappa}X^{\epsilon}_{(i)}\partial_{\lambda}X^{\zeta}_{(i)}\Bigr) \ .
\label{eq:5daction}
\end{eqnarray}
Here $\phi$ the is the  modulus of the Calabi-Yau (CY) threefold and
${\cal A}_{\gamma\delta\epsilon\zeta}$ is a 4-form gauge field. According
to \cite{KOST},  the left brane is the visible one and it is assigned a
negative tension $-\alpha=\alpha_3$, the bulk brane has positive tension
$\beta=\alpha_2$ and the hidden brane has a positive tension
$\alpha-\beta= \alpha_1$. The solution  for the 5-form was written as follows:
\begin{equation}
  {\cal F}_{0123y}=\alpha D^{-2}\quad  {\rm for}\qquad y<Y \, ; \qquad
  {\cal F}_{0123y}=(\alpha-\beta)
  D^{-2} \quad {\rm for}
   \qquad  y>Y
\label{F}
\end{equation}

This action with three branes was not derived from supersymmetric theory.
The action with two branes, which was derived in
\cite{lukas,universe,kelly} from Ho\u{r}ava-Witten (HW) theory, does not
have a 4-form ${\cal A}_{\gamma\delta\epsilon\zeta}$, either in the bulk
or on the branes. The appearance  and the role of the 4-form gauge field
in  the five-dimensional supersymmetric bulk \& brane action was explained
in \cite{BKV}, but not in the context of HW theory.  
Recently it was shown in  \cite{KKLT} that one can actually derive the action of the type of (\ref{eq:5daction}), but the factor in the action in front of ${\cal F}^2$ has to be corrected so that the action
correspond to a bosonic part of the supersymmetric  action.
Also the factor and the sign in (\ref{F}) have to be changed. It is easy to verify that the WZ term in the brane action does not cancel the BI term for the `solution'
given in \cite{KOST}, which proves that it is not a BPS solution  before the  correction is made. Thus one can find a BPS solution with the boundary and bulk brane present, see \cite{KKLT}, but it is somewhat different from the one presented in \cite{KOST}.

Note that the  BPS property of the classical solution is not the same as the requirement of unbroken supersymmetry. If the unbroken supersymmetry is established, the solution always has a BPS feature: the energy  
takes its minimal value
and the supersymmetry bound is saturated, see e.g. \cite{Kallosh:1992ii}. However,  non-supersymmetric BPS configurations are also possible. The difference is that for supersymmetric solutions one may expect that they will remain BPS states even with an account taken of quantum corrections,  because of non-renormalization theorems for supersymmetric BPS states \cite{Kallosh:1992gu}. Meanwhile, non-supersymmetric BPS solutions may loose their BPS properties  
when quantum corrections are taken into account.

The issue of the unbroken supersymmetry for multi-domain wall
solutions is more complicated than for other multi-brane solutions with
co-dimension greater than 2, like multi-black-holes, multi-3-branes etc.
The major difference is the behavior of the form fields at large
distances. The supersymmetric domain walls are charged and the form fields
are constant. They exist therefore only in the compact space with the
vanishing total charge. The explicitly supersymmetric solution requires
apart from supergravity the presence of the supersymmetric source actions
which take care of the jump conditions on the wall.

The status of unbroken supersymmetry of the  solution with two boundary
branes without a bulk brane may be inferred from an action in
\cite{universe} and~\cite{kelly} where the Born-Infeld part of the brane
actions at the fixed points of the orbifold  was given. However when the
bulk brane is present in addition to boundary branes, the issue of unbroken
supersymmetry is less clear. The
supersymmetric bulk \& brane construction of \cite{BKV} (and HW theory)
allows one to prove clearly the unbroken supersymmetry only for the case
that the positive and negative tension branes are placed at the fixed
points $0$ and $R$ of the orbifold ${S^1\over {\bf Z}_2}$. When the brane
is not at the fixed points, the supersymmetry variation of the 4-form
field in the Wess-Zumino term of the brane source action is not
compensated by the supersymmetry variation of the Born-Infeld term. This
makes the unbroken supersymmetry of the multi-domain walls problematic
despite the fact that the bosonic solution with the jump of the 5-form
field strength can be given.

Analogous problem exists for the multi-domain walls in D8-O8-system of
type IIA string theory \cite{D8O8}. The variation of the R-R 9-form in the
Wess-Zumino terms of the brane action depends on the NS-NS 2-form
$B_{\mu\nu}$, whereas the supersymmetry variation of the Born-Infeld term
does not depend on $B_{\mu\nu}$. Therefore the BI+WZ source action at the
fixed points of the orientifolds is supersymmetric due to the fact that
the NS-NS 2-form $B_{\mu\nu}$ is odd under ${\bf Z}_2$-symmetry and
vanishes at the fixed points, but not between them. Therefore only when
all D8 domain walls are coincident with orientifold planes, the
supersymmetric bulk\&brane action is available and the unbroken
supersymmetry of the solution can be proven in \cite{D8O8}. The situation
with the supersymmetry of the multi-wall solutions remains an open issue.

\subsection{Tension on visible brane}

Let us now compare  the ekpyrotic  construction with HW theory.  At the
time when HW theory was suggested the issue of brane tension was not
emphasized. More recently in the Randall-Sundrum I scenario \cite{RSI} the
positive  and negative  tension branes were given the  names, `hidden' and
`visible' brane, respectively. In this version the  hidden brane was at
the left, at $y=0$, and the visible brane at the right, at $y= \pi r_c$.
The `visible' brane, called sometimes the ``TeV brane'', was designed to
provide a solution of the hierarchy problem due to the fact that the warp
factor decreases exponentially towards the `visible' brane. In RS II
\cite{RSII} a dramatic change of the setup was made as the labels of
`hidden' and  `visible' branes were reversed. The former 'visible' became
hidden,  the former `hidden' became visible and it was suggested to send
the negative tension hidden brane  out of the world.

With all this in mind we will consider the known facts about HW and
heterotic M-theory and the tensions of various  branes in
agreement with supersymmetry. 

\subsubsection{Standard embedding}

In case of boundary branes of the original
HW theory a standard embedding corresponds to a positive tension visible
brane and negative tension hidden brane. Let us shortly remind how this happens.
We will  look at the heterotic M-theory in \cite{kelly} where the Born-Infeld part of the brane actions were derived
from HW theory. The  tension of the visible brane is given by
$-\alpha_{vis}$, where (we ignore here some irrelevant positive constants)
$$
-\alpha_{vis} =  {1\over 8\pi^2} \int _C   tr R \wedge R =  n \ ,
$$
see for example eq. (3.13) in \cite{kelly}. Here the integration is over a
supersymmetric cycle  of the CY manifold, and $R$ is the curvature form of
the internal manifold.  The integer $n$ characterizes the first Pontrjagin
class of CY. Thus the tension of the visible brane
 is positive. 
 
 Let us present here a few important  steps of
the derivation of the tension formula. The
tension on each brane, according to \cite{kelly}, is proportional to
$
-\alpha_{vis}= {1\over 4\pi^2} \int _C  ( tr F^{vis}\wedge F^{vis}-
{1\over 2} tr R\wedge R) \ ,
$ and 
$
 -\alpha_{hid} = {1\over 4\pi^2} \int _C  (tr F^{hid}\wedge F^{hid}-
{1\over 2} tr R\wedge R) $.

So far the visible and hidden brane are treated on an equal footing. The
difference is in spin embedding in which only the visible brane
participates. On the visible brane a standard spin connection embedding
was performed so that the background is
$
  tr F^{vis}\wedge F^{vis}=
tr R\wedge R, \,  F^{hid}=0,
$
which implies that
$
tr (F^{vis}\wedge F^{vis}+ F^{hid}\wedge F^{hid})- tr R\wedge R = 0 \ .
$.
The $E_8$ gauge theory on the visible brane was broken to its subgroup
$SU(3)\times E_6$ and after spin embedding  on the visible brane there are
$E_6$ gauge field excitations and on the hidden brane there are  $E_8$
gauge field excitations. On the visible brane, where spin embedding takes
place and  $tr F^{vis}\wedge F^{vis}=
 tr R\wedge R$, we have
\begin{equation}
-\alpha_{vis} =  {1\over 8\pi^2} \int _C  tr R\wedge R >0\ ,
\label{tensionV}
\end{equation}
on the hidden one with $F^{hid}=0$ the tension is
\begin{equation}
-\alpha_{hid}=  - {1\over 8\pi^2} \int _C tr R\wedge R  <0\ .
\label{tensionH}
\end{equation}
This  explains  how the visible brane in the original HW
theory with standard embedding acquires a positive tension.

In \cite{Witten} it was explained by
Witten that the volume of the CY space $e^{\phi}=V$ at the visible brane
at $y=0$ is larger than that at the hidden brane  at $y= R$:
$ V(0) > V(R) $ and the gauge coupling on the hidden brane  is stronger
than the one on the visible brane, $g^2_v < g^2_h$ due to the
inverse relation between the CY volume and the gauge coupling: 
\begin{equation}\label{coupling}
 { g^2_v \over g^2_h} \sim { V_h \over V_v} \sim {D^3(R)\over D^3(0)}\ .
\end{equation}
The subsequent
work on HW phenomenology \cite{HWph} is based on a strong coupling at the hidden brane
required for gaugino condensation on
the hidden brane.

\subsubsection{Non-standard embedding}

The presence of  the bulk brane requires using the  non-standard embedding. The total tension (and
the total charge) of all branes must vanish. In HW case  the tension
of the visible brane was positive and opposite to the tension of the hidden
brane:
$\alpha_v= -\alpha_h>0$. Now the new relation between brane tensions is
\begin{equation}
  \alpha_v + \alpha_h+ \beta =0 \ .
\label{tensions}
\end{equation}
This correspond to a cohomology constraint
\begin{equation}
c_2(V_1)+ c_2(V_2) - c_2(TX)+[W]=0 \ ,
 \label{cohomology}
\end{equation}
where $[W]$ is the cohomology class associated with the five-branes and
$c_2(V_1), c_2(V_2),  c_2(TX)$ are the second Chern class of the gauge
bundles $V_1, V_2$ and of the tangent bundle $TX$, respectively.

The situation with the tension in non-standard embedding is the following. If one adds the bulk brane as a small modification of the previous two-brane configuration, one still has the visible brane with positive tension. However, in general  
one can have examples of both positive and negative tension on visible brane  \cite{Benakli:1999sy,Donagi:2001fs}. 

In the ekpyrotic scenario the ratio between the tensions of boundary and
bulk branes was chosen to be extremely small, ${\beta\over \alpha}=4\cdot
10^{-4}$.
No  examples with negative tension visible brane and ${\beta\over \alpha} \ll 1$ have been considered in the
literature until very  recently   \cite{Donagi:2001fs}. The  examples considered in \cite{Donagi:2001fs} require a special assumption that the volume of the base curves is much larger than the volume of the fiber curve in CY space. But even for such examples one still has an additional problem, which we are going to explain now.

The switch to the non-standard embedding does not change the relation between the volume of the Calabi-Yau space and the gauge coupling, Eq. (\ref{coupling}).
When the visible brane tension is positive (the warp factor decreases towards
the negative tension brane), the standard HW phenomenology applies. One can have small gauge coupling ${g^2_{GUT}\over 4\pi} \sim 0.04$ on the visible brane, and large coupling ${g^2_{hidden}\over 4\pi} = O(1)$ on the hidden brane, which may lead to gaugino condensation \cite{Witten,HWph,lukas}.

However, when the visible brane tension is negative, a substantial
modification of the standard HW phenomenology is required since in this case the gauge coupling constant on the hidden brane is
smaller than on the visible brane.  For example, with the parameters of \cite{KOST} one has ${g^2_{hidden}\over 4\pi} \sim 10^{-3}$ on the hidden brane, which is way too small for the gaugino condensation.  In general, one may try to develop acceptable phenomenology  with stronger coupling observable sector \cite{Benakli:1999sy}, but this unconventional possibility is
much less understood and developed that the standard HW phenomenology
\cite{Witten,HWph,lukas}.

\subsubsection{Tensions and BPS harmonic functions}

The relation between brane tensions and  the volume of CY manifold
(inverse to the square of the gauge coupling) can be seen in the BPS solution where the sign of the tension on the visible brane defines the slope of the CY
volume. The volume of the CY space is
given by a harmonic function
\begin{equation}
  e^{\phi}= B D^3 \ , \qquad D= C + \alpha y , \qquad y <Y \ ,
\label{volume}
\end{equation}
in  notation of \cite{KOST}. In \cite{kelly}, where the standard embedding was used,
$C>0$, but the sign of $\alpha$ was not discussed. If it were noticed in
\cite{kelly} that $\alpha$ in this equation is negative\footnote{ It was
confirmed to us by D. Waldram in private discussion that indeed $\alpha$
in the harmonic function in equation  (5.9) in \cite{kelly} must be
understood as negative.} one would be forced to rewrite the harmonic
function as follows:
\begin{equation}
  \tilde D= C - |\alpha| y \ , \qquad \alpha<0 \ ,
\label{volumeCorr}
\end{equation}
and comment on the existence of the critical distance between walls
\begin{equation}
  \tilde D(y_{crit})= 0 \qquad \Rightarrow  \qquad y_{crit} =
  {C\over |\alpha|}  \ .
  \label{crit}
\end{equation}
This would prompt a requirement that the second wall has to cut off the
singularity of the space time-metric when $\tilde D=0$ and therefore
\begin{equation}
  R< y_{crit}\ .
\label{restriction}
\end{equation}
Precisely this situation occurs in many cases of supersymmetric domain
walls in \cite{BKV,D8O8} where always the warp factor falls down away from
the visible positive tension brane and one has to take care of the maximal
distance between the walls.

However it was not noticed that the tension of the visible brane is positive in
\cite{kelly}, where the standard embedding was used and the harmonic
function was taken in the form $D= C + \alpha y$ instead of $\tilde D= C - |\alpha|
y$. The authors of \cite{KOST} used the same notation as in \cite{kelly} and  assumed, as equation $D= C + \alpha y$ suggests, that  $\alpha$ is positive, i.e. the brane tension  is negative,  and $D(y)$ decreases near the visible brane at $y = 0$. This assumption, combined with  the idea that the density perturbations are produced in their scenario because of the decrease of $D(y)$ at small $y$, has led to the conclusion that it is  ``necessary  for the visible brane to be in the small-volume region of space-time''  \cite{KOST}.

However, as we will show in Sect. \ref{perturb}, this requirement is not necessary. We do not see any obvious reason to use unconventional versions of the HW theory and insist that we must live on the negative tension brane. 

On the other hand, one cannot improve the situation by flipping the sign of $\alpha$ in all expressions in \cite{KOST}. Indeed, according to the original version of the ekpyrotic scenario,  $C=100$, and $\alpha = 250 M_5$. If we simply change the sign of
$\alpha$ and take $\alpha = -250 M_5$ we will find that the singularity of
the metric appears between the walls at $y=0$ and at $y=R= M_5^{-1}$:
\begin{eqnarray}
  &&\tilde D =  C - |\alpha| y = 100- 250 M_5 y \ , \nonumber \\
  &&y_{crit} =  {2\over 5}M_5^{-1}< R= M_5^{-1} \ .
\label{corr}
\end{eqnarray}
A quick fix of this problem is possible: one can change  the parameters,
e.g. by making  $C$ larger or $|\alpha |$ smaller so that $y_{crit}=
{C\over |\alpha|}> R$ and the naked singularity at $y_{crit}$ is cut off
by the hidden wall.

\begin{figure}[b]
\centering\leavevmode\epsfysize=8cm \epsfbox{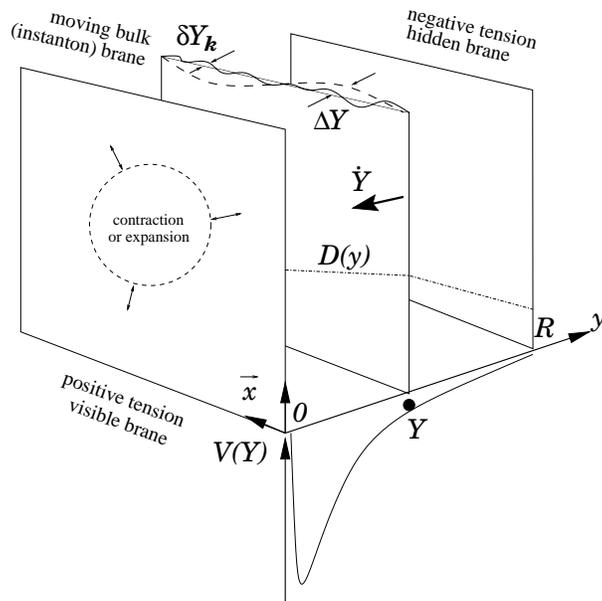}
\caption[fig2]{\label{fig2} Sketch of the pyrotechnic  scenario. We live
on a brane with {  positive} energy density. The volume of space
controlled by the metric $D(Y)$ decreases near our brane, which gives a
red tilt to the spectrum of perturbations. The mechanism for the
generation of fluctuations  $\delta Y_k$ in this scenario, as well as in
the ekpyrotic  scenario, amplifies all inhomogeneities, including
classical inhomogeneities $\Delta Y$ of the bulk brane. }
\end{figure}

In what follows we will call the improved  model the `pyrotechnic
universe,'  see Fig. \ref{fig2}, where a  sketch of the properties of the
model is given. This model will have many of the same problems as the ekpyrotic scenario, but it has two advantages. First of all, it is based on the conventional HW phenomenology. Second advantage is that we are not going to insist that this model solves all cosmological problems without using inflation. As we will see, it is very hard or even impossible to do so. Moreover, avoiding inflation requires additional fine-tuning.  In the ekpyrotic scenario one should deviate from the usual HW phenomenology and give up all advantages of inflationary theory. We do not see any reason to do it.

\section{A new mechanism for the generation of density perturbations with any
kind of spectrum\ldots and why it might not work}\label{perturb}

\subsection{A simple 4d example}

In this section we will consider the mechanism for the generation of
density perturbations in the ekpyrotic and pyrotechnic scenarios. Instead
of considering a complicated setting with three branes moving in an
expanding  five-dimensional universe, let us first consider a simple
problem of motion of a scalar field falling down from the top of the
effective potential $V(\phi)$ with $V'' < 0$ in  four-dimensional
Minkowski space. As we will see, these two problems are directly related
to each other.

In what follows we will consider two particular examples: $V = -V_0 e^{-
\phi/M}$ and $V = -{\lambda_n} \phi^n/M^{n-4}$ with $n > 2$; $M$ is some
constant of dimension of mass. In general, one needs to add to these
potentials some other terms that stabilize the motion of the scalar field
after it falls down and ensure that the effective potential vanishes at
its minimum. We will return to this important point later.

In both cases the curvature of the potential is negative, $V''(\phi) <0$,
and its absolute value grows when the field falls down to smaller values
of $V(\phi)$. Conservation of energy implies that ${\dot\phi^2\over 2} +
V(\phi) = E$, where $E$ is some constant. We will assume for simplicity
that the field was falling from the top of the potential with vanishing
initial energy, $E = 0$, so that
\begin{equation}\label{velocity}
 |\dot\phi| = \sqrt{-2 V(\phi)} \ .
\end{equation}
Quantum fluctuations of the scalar field  $\delta \phi_k(t) e^{i{\vec k
\cdot \vec x}}$ living in a background of a homogeneous field $\phi(t)$
satisfy the equation
\begin{equation}\label{tach}
\ddot{\delta\phi_k}+ \left(k^2 + m^2(\phi) \right)\delta\phi_k=0  \ ,
\end{equation}
where  $m^2(\phi) = |V''(\phi)|$. If $V'' <0$ the modes with $k^2 <
|V''(\phi)|$ do not oscillate. Instead, they grow exponentially. For
example, if $V'' = -m^2 = const$, one has $\delta\phi_k \sim \exp{\sqrt{
m^2 - k^2 }\ t}$. The initial amplitude of the growing modes depends on
the initial conditions. Since the initial curvature of the potential is
very small, one may assume that the initial amplitude of fluctuations is
the same as in the theory of  a massless scalar field, so that $ \langle
\delta\phi^2 \rangle
 =  \int {  dk^2  \over 8\pi^2 }$. Ignoring the coefficients $2$ and $\pi$,
one may say that the average amplitude of fluctuations  with momenta $\sim
k$ is proportional to $k$: $\delta \phi(k) \sim k$. (This is analogous to
the famous relation $\delta\phi(k) \sim H/2\pi$ during inflation.) At
large $t$ this amplitude grows as $\delta\phi(k) \sim k \exp{\sqrt{ m^2 -
k^2 }t}$. Exponential growth of $\delta\phi (k)$ can be interpreted as
generation of a classical field $\delta\phi(k)$.  This is the basic
feature of the theory of tachyonic preheating developed recently in
\cite{tach} in a different context.

In the models such as $V = -V_0 e^{- \phi/M}$ and $V = -{\lambda_n }
\phi^n/M^{n-4}$ with $n > 2$, the curvature of the effective potential
grows while the field falls down from the top of the potential. Therefore,
in the beginning  only the modes with extremely small $k$ are growing
exponentially, whereas short wavelength perturbations with $k >
\sqrt{|V''|}$ are oscillating with a nearly constant amplitude. However,
when the field $\phi$ grows, the value of  $|V''|$ grows too, and new
modes with momenta $k < |V''|$ stop their oscillations and start growing,
with the initial amplitude $\delta \phi (k) \sim {k} \sim {\sqrt
{|V''(\phi)|}}$ \cite{tach}.

These fluctuations change the local value of the field $\phi$ and
therefore they lead to a delay of the moment when the field $\phi$ rolls
down to some value $\phi_0$, which  corresponds to the minimum of
$V(\phi)$ where the process of reheating begins (or to the brane collision
in the five-dimensional setting  if  $\phi$ is identified with the modulus
$Y$). A detailed discussion of the process of the growth of fluctuations
will be given in the Appendix. Here we will give a simple shortcut to the
answer.

To calculate the delay of reheating produced by these fluctuations one
should divide $\delta\phi$ by $|\dot \phi|$. This can be done at any time
after $\delta\phi_k$ stops oscillating. Indeed, for small $k$ the
equations for $\delta\phi$ and $|\dot \phi|$ coincide, so their ratio
remains constant. Suppose that we transplant this picture into an
expanding universe with the value of the Hubble constant $H$ induced by
the field $\phi$. Then for the wavelength $\lambda \sim k^{-1}$ greater
than the size of the horizon, the time delay  $\delta t_k$,   according to
\cite{pert}, results in adiabatic density perturbations
\begin{equation}\label{pert}
{|\delta_k|} \sim   H \delta t_k \sim H {\delta\phi(k) \over\dot\phi} \sim
H \cdot {\sqrt {|V''(\phi)|}  \over \sqrt{ |V(\phi)|}} \ .
\end{equation}
The numerical coefficient in these equations depends on various details
such as the equation of state of the universe, but typically it is of
order  unity, so we will omit it in what follows.

For the exponential potential $V = -V_0 e^{- \phi/M}$ one finds density
perturbations
\begin{equation}\label{pertex}
{|\delta_k|} \sim  {H  \over  M}   \ .
\end{equation}
Note that this amplitude  does not depend on $k$, i.e. these perturbations
have a {\it flat}  spectrum, $n_s=1$, like in inflation, but without any
inflation! To obtain this result we did not need any brane physics or
string theory, it is a trivial consequence of  the tachyonic instability.

For the power-law potential $V = -{\lambda_n } \phi^n/M^{n-4}$ we find
\begin{equation}\label{pertpow}
{|\delta_k|} \sim  \sqrt{n(n-1)}~  {H \over  \phi}   \ ,
\end{equation}
or, in terms of $k \sim \sqrt {|V''(\phi)|}$,
\begin{equation}\label{pertpow2}
{|\delta_k|} \sim (n(n-1))^{n\over 2(n-2) }~  {H \over  M^{n-4\over n-2 }
} ~ k^{-{2\over n-2}} \ .
\end{equation}
 The spectral index is given by
\begin{equation}\label{index}
n_s= 1+{d\log|{\delta_k}|^2\over d \log k} = 1- {{4\over n-2}} <1 \ .
\end{equation}
 We see that the spectrum in the class of power-law  potentials  is always
{\it red}, $n_s <1$. For instance, for $n=3,4,5$ we have $n_s=-3, -1, -1/3
$, respectively. All of these spectra are observationally unacceptable
(too red). To have $|n_s-1| <  0.1$, as suggested by cosmological
observations \cite{Bond}, we must have $n >  40$.

These results are valid for the power-law potentials with negative $n$ as
well, in which case the spectrum becomes {\it blue}. For example, for $V =
-{\lambda_n } M^{4-n}/\phi^n$ one has
\begin{equation}\label{index2}
n_s= 1+{d\log|{\delta_k}|^2\over d \log k} = 1+ {{4\over n+2}} <1 \ .
\end{equation}
Once again, to have $|n_s-1| <  0.1$  we must have $n >  40$. Thus, one
should make the very special choice of a nearly exponential potential  to
produce a cosmologically acceptable spectrum.

In general,   the ``color'' of the spectrum  depends on the way
$|V''(\phi)|$ behaves with respect to $|V(\phi)|$ when the field $\phi$
rolls down to the minimum of $V(\phi)$. For example, the potential $V =
-V_0 e^{- \phi^2/M^2}$ would lead to a {\it blue} spectrum of density
perturbations, decreasing as $\log^{1/2} k$ at small $k$.

\subsection{The same mechanism in 5d brane cosmology}

The reason we discussed this mechanism here is that it provides a simple
interpretation and generalization of the mechanism of production of
density perturbations in the ekpyrotic and pyrotechnic scenarios
\cite{KOST}. The discussion of this effect in \cite{KOST} is very involved
because the authors were trying to follow simultaneously the expansion of
the universe, motion of all three branes and perturbations of the bulk
brane.  At first glance it seems to be an extremely complicated
gravitational problem. To treat it properly the authors introduced the
effective scale factor which was not a real scale factor but in fact
something very much different, and as a result the physical meaning of
this effect became rather difficult to analyse.

However, the motion of all branes and the total change of metric during
the whole duration of the process of motion of the bulk brane towards the
visible brane is rather insignificant. As we will see, in order to
understand the mechanism for the generation of density perturbations in
the first approximation,  one can completely neglect expansion of the
universe. In this case, the equation of motion for the brane at a distance
$Y(x)$ from the visible brane is completely analogous to the equation for
the scalar field discussed above, and it becomes clear that the mechanism
of generation of density perturbation in the ekpyrotic scenario is exactly
equivalent to the effect of tachyonic instability described in the
previous subsection.

Indeed, density perturbations discussed in \cite{KOST} are produced due to
the fluctuations of the bulk brane. According to \cite{KOST}, the
effective Lagrangian of the brane in the lowest approximation in
${\beta\over \alpha}$ is given by
\begin{equation}\label{efflag}
{\cal L}_{\beta}= 3\beta M_5^3 B\left[\frac{1}{2}
D(Y)^2\eta^{\mu\nu}\partial_{\mu}Y\partial_{\nu}Y -V(Y)\right].
\end{equation}
This Lagrangian looks like a Lagrangian of a scalar field $Y(x)$ with a
nonminimal kinetic term. To properly normalize $Y$ (ignoring for a moment
the insignificant dependence of $D(Y)$ on $Y$ at $Y \approx R$) one should
multiply $Y$ by $\sqrt{3\beta M_5^3 B}$, i.e. introduce the variable
\begin{equation}\label{phi}
\phi(x) = \sqrt{3\beta M_5^3 B}\, D(Y)\, Y(x) .
\end{equation}
This is necessary in order to find the correct amplitude of quantum
fluctuations of $Y$. A properly normalized  potential is $ {\cal V}(\phi)
= 3\beta M_5^3 B\, V(Y)$ and a properly normalized term ${\cal V}''(\phi)
= {d^2 {\cal V}\over d \phi^2}$  is ${d^2 V\over d Y^2} D^{-2}(Y)$.
 So the properly normalized amplitude of quantum perturbations of the field
$\phi$ at the moment when it stops oscillating
 is $\sqrt{|{d^2 V\over d Y^2}|} D^{-1}(Y)$. (For simplicity, here and in the
following equations we are ignoring factors of 2, 3 and $\pi$.)

Returning to the amplitude of fluctuations of $Y$, one finds
\begin{eqnarray}\label{Yk}
\delta Y(k) &\sim & {\delta \phi\over \sqrt{\beta M_5^3 B}} \sim {\sqrt{|
{\cal V}''|} \over  \sqrt{\beta M_5^3 B}\, D(Y)  }  \nonumber \\
 &\sim &  {k  \over
\sqrt{\beta M_5^3 B}\, D(Y) }  \sim { \sqrt{|{d^2 V\over d Y^2}|}  \over
\sqrt{\beta M_5^3 B}\, D^2(Y) } .
\end{eqnarray}
Dividing this result by $\dot Y = D^{-1}(Y) \sqrt { |V(Y)|}$ one finds the
time delay
\begin{equation}
\delta t(k) \sim   { \sqrt{|{d^2 V\over d Y^2}|}  \over  \sqrt{\beta M_5^3
B\ V(Y)}\, D(Y)  } .
\end{equation}

For the exponential potential $V(Y) = -v e^{-\alpha m Y}$ one has
\begin{equation}
\delta Y(k) \sim   {m\alpha \sqrt { v e^{-\alpha m Y }}  \over
\sqrt{\beta M_5^3 B}\, D^2(Y) } ,
\end{equation}
and the time delay
\begin{equation}
\delta t(k) \sim   {m\alpha \over  \sqrt{\beta M_5^3 B}\, D(Y)  } .
\end{equation}
This result coincides with the result obtained in Eq. (72) of Ref.
\cite{KOST} up to a factor of $1/\sqrt {3B} \sim 20$.\footnote{The reason of this disagreement, which affects the final answer for the amplitude of density perturbations, is that the authors of
\cite{KOST} did not include $3B$ in the normalization of $Y$.}
After multiplying this result by $H$, we again obtain a nearly flat
spectrum. It is slightly blue if $D(Y)$ decreases towards small $Y$, as
assumed in \cite{KOST}, and it is slightly red if $D(Y)$ increases towards
small $Y$ as in  the pyrotechnics scenario. But the  mechanism of
generation of perturbations works (or does not work)  independently of the
slow decrease or slow increase of $D(Y)$, and the color of the spectrum
often is much more sensitive to the choice of the potential $V(Y)$ rather
than to the behavior of $D(Y)$. In particular, for any power-law potential
$V(Y) \sim -Y^n$ one would get an unacceptably red spectrum unless $n >
40$. The red tilt introduced by the growth of $D(Y)$ at small $Y$ implies
that $n$ must be even greater. In this respect the ekpyrotic/pyrotechnic
scenario is much less robust than the inflationary universe scenario where
the spectrum is nearly flat for almost all inflationary models.

\subsection{Is this a realistic mechanism?}

It could seem that now we have a new realistic mechanism for the
generation of density perturbations, which does not require inflation, and
which can produce perturbations with any spectrum we like, depending on
the choice of $V(Y)$ or $V(Y)$. As we have seen, there is nothing
brane-specific in this mechanism. Explanation of this mechanism required
nothing but two simple equations, (\ref{pert}) and (\ref{pertex}). So why
did we not use this mechanism before if it is so trivial? Is there a
catch?

There are two different problems related to it. First of all, we needed to
{\it assume} that the universe was flat and homogeneous from the very
beginning. Of course, one may argue that our universe initially {\it must}
be flat and homogeneous for some reason to be discovered later. This was
the ideology of the models of structure formation due to topological
defects or textures, which sometimes were advertised as the models that
``match the explanatory triumphs of inflation while rectifying its major
failings'' \cite{SperTur}. In our opinion, if we  find that inflationary
theory does not work, we may use such models as a ``plan B," but we will
definitely loose a lot by doing so.

A more serious problem is that in our discussion of fluctuations in 4d we
neglected expansion of the universe induced by the effective potential
(i.e. the term $3H\dot{\delta \phi_k}$ in eq. (\ref{tach})). Therefore,
our results apply only for the  short wavelength fluctuations with $k^2
\sim |V''| > H^2$, where $H$ is the Hubble constant. This means that all
perturbations that are correctly described by this method must have
wavelength $k^{-1}$ smaller than $H^{-1}$. Thus, these perturbations are
of no interest for the theory of the large-scale structure formation
unless one makes some trick to make $H$ exponentially small during  the
process of generation of the perturbations. This does not mean that no
perturbations are produced with wavelength larger than $H^{-1}$. At the
stage when $|V''| \ll H^2$, the universe experiences inflation, so we get
usual inflationary perturbations. A good thing about it is that such
perturbations have a flat spectrum for a much broader range of potentials
such as $-\phi^n$ with  $n < 40$, as well as with $n > 40$. Thus we are
back to inflationary theory.

Nevertheless,  if our goal is to avoid using anything inflationary, we may
still try to do so. For example, we can avoid any gravitational
backreaction if we assume that the effective potential  vanishes near its
maximum,  so that it does not induce any Hubble constant during the first
part of the process. This was exactly the assumption made in \cite{KOST}.
They considered the potential $V(Y) = -v \exp ({-\alpha m Y})$ which
nearly vanishes at $Y = R$. Therefore, the Hubble constant initially also
vanishes, and the new mechanism for the generation of density
perturbations works. However, the price one may pay for it is that after
the field
 $\phi({\vec x})$ (or $Y({\vec x})$) rolls down to the minimum of the effective
potential and the energy of its oscillations dissipates, the effective
potential remains large and negative, and we may find ourselves in a
universe with a large negative cosmological constant. The authors of
\cite{KOST} are aware of this problem but  argue that it can be somehow
resolved, assuming  that $V(Y)$ may suddenly rise to zero at $Y=0$ due to
some nonperturbative effects. Until this problem is resolved, the
existence of a novel realistic mechanism for the generation of density
perturbations with a flat spectrum remains an interesting but speculative
possibility.

Of course this issue would not arise if we assume that initially $V(Y)$ is
positive, and then eventually the bulk brane falls to $Y=0$ with $V(Y) =
0$. But in this case initially we will have inflation, which will produce
inflationary perturbations with flat spectrum. Thus, having inflation as a
part of brane cosmology may not be a bad idea after all.

Let us check how easy it would be to avoid inflation while still producing
density perturbations with a flat spectrum on a scale comparable to the
observable part of the universe, $l \sim 10^{28}$ cm. At the beginning of
the big bang (brane collision) our part of the universe was smaller by a
factor of $T/T_0$, where $T \sim 10^{11}$ GeV in the reheating temperature
in the ekpyrotic scenario  \cite{KOST} and $T_0$ is the present
temperature $\sim 3 K$. This ratio is about $10^{24}$. This means that the
wavelength of the perturbations we are discussing  was $k_0^{-1} \sim
10^4$ cm at the moment of the brane collision, so that
\begin{equation}\label{k0}
k_0 \sim 10^{-17} {\rm GeV}  \sim 10^{-32} M_5 \ .
\end{equation}

To produce perturbations on this scale by the tachyonic instability rather
than due to inflation one needs to have $H < k_0$. According to Eq. (20)
of \cite{KOST},
\begin{equation} \label{flat}
H^2 = \left({\dot a \over a}\right)^2 =  b\left({1\over 2} D^2\dot Y^2 +
V(Y)\right),
\end{equation}
where $b \sim 10^{-14} M_5^2$. The authors of  \cite{KOST} have chosen
$V(R)$ to be nearly zero, $V(R) = -v e^{-\alpha m R} \sim - 10^{-120}$,
whereas at small $Y$ one has $V(Y) \sim - 10^{-10}$. This negative
potential has not been really derived from a fundamental theory, and, as
it was argued in \cite{DvaliTye}, where a similar scenario was developed
as a basis for inflationary theory, this potential may contain many other
terms of different nature. One of the problems with this potential becomes
obvious if one tries to assume that the bulk brane initially did not move,
$\dot Y=0$, which looks like a very natural assumption. Then the equation
$H^2 \sim  b({1\over 2} D^2\dot Y^2 + V(Y))$ becomes inconsistent for
negative $V$. This is an indication that either the bulk brane has a more
complicated geometry, like an open universe created by tunneling, or one
should add some positive term to $V(Y)$. So let us see  whether anything
will change if we add to it a very small positive constant $V_\Lambda$
such that $|V(R)|\ll V_\Lambda \ll  |v|$.

In such a case, the later stages of the bulk brane motion will not change
at all, but in the beginning of its evolution it will experience inflation
with $H^2 > 10^{-14} M_5^2 V_\Lambda$. This will be similar in spirit to
the Dvali-Tye scenario \cite{DvaliTye}. Inflation will induce the usual
inflationary perturbations with flat spectrum on a scale $k_0  \sim
10^{-32} M_5$ unless one fine-tunes $V_\Lambda$ to be incredibly small,
$V_\Lambda < 10^{-50}$.

To summarize, if one wants to use a tachyonic instability to produce
density perturbations with a flat spectrum, one must fine-tune the
functional form of the potential. For example, one should avoid such
potentials as $ -Y^{\pm n}$ with $n < 40$. Then one must solve the problem
of the negative cosmological constant at the end of the process and
fine-tune the value of $V(Y)$ near the hidden brane with accuracy
$10^{-50}$ in the natural units of $M_5^4$. This last step  should be made
if one wants to  avoid using a much more robust method of generation of
density perturbations  provided by inflation.

\section{Problems with the bulk brane potential $V(Y)$.}\label{pot}

 One of the most crucial assumptions of the ekpyrotic scenario is
the existence of the exponential potential of the bulk brane,
\begin{equation}\label{1pot}
V(Y) \sim -v e^{-|\alpha m| Y} \ .
\end{equation}
This potential was added to the model by hand. It was also necessary to
assume that this potential is not purely exponential, but it rises to zero
at $Y=0$. But does this potential correctly describe the situation, or
something else should be added to it? This is a very important issue
because this potential  is of the order $10^{-120}$ near the hidden brane \cite{KOST},
so one must avoid any corrections to this potential with an accuracy
$10^{-120}$  to keep the scenario intact.

If one adds an exponentially small positive constant to $V(Y)$, one gets inflation,
as  in \cite{DvaliTye}. Such terms  as $Y^{\pm n}$  with  $ n  < 40$ also
should be forbidden. If the brane configuration is a supersymmetric BPS
state, one may assume that such terms cancel each other and vanish.
However, as we already mentioned, supersymmetry of the three-brane
configuration in HW setting is not rigorously established.  Moreover,
there is no supersymmetry in the real world, so the cancellation of the
long-range forces cannot be exact. Can we really suppress power-law terms
with accuracy $10^{-120}$? Also, even if these terms were absent for
exactly parallel branes, they would appear again if the branes are not
exactly parallel, or if they become non-parallel because of the quantum
fluctuations $\delta Y$ produced due to the tachyonic instability
\cite{Dvali}. Indeed, if the branes are non-parallel, they are not in a
BPS state, and the cancellation of the long-range forces acting between
the branes is again not exact, see e.g. \cite{Ts,DvSh}. Meanwhile we need
it to be exact with accuracy $10^{-120}$.

There are other issues to consider as well. In the   ekpyrotic scenario
the positive tension hidden brane splits into two positive tension branes.
As we have shown, however, this setting contradicts the usual HW
phenomenology. If one makes the standard assumption that the hidden brane in HW
scenario has negative tension, does it mean that it splits into two
negative-tension branes? It is very hard to imagine that such  a process
is possible. But it is equally hard to imagine that it splits into a
positive tension bulk brane and a negative tension hidden brane with an
{\it increased} absolute value of tension. Do we  have a run-away brane
instability where the brane tension tends to become indefinitely large?

Let us return now to a much simpler and less ambiguous issue and analyse
the assumption made in \cite{KOST} that the energy of  the  bulk brane is
negative and is given by $V(Y) \sim - v e^{-|\alpha m| Y}$. According to
\cite{KOST}, such Yukawa-type terms may appear for precisely parallel
branes because of non-perturbative effects such as the exchange of virtual
M2-branes  between the bulk brane and either of the boundary branes.
However, as  shown by Moore, Peradze and  Saulina \cite{Moore:2000fs}, the
structure of the potential can be much more complicated.

Before discussing the general structure of the potential obtained in
\cite{Moore:2000fs}, let us first discuss the  mechanism outlined in
\cite{KOST} that could lead to the potentials such as $V(Y) \sim - v
e^{-|\alpha m| Y}$.

First of all, let us represent this potential as a function of $\phi(x) =
\sqrt{3\beta M_5^3 B}\, D(Y)\, Y(x)$, as we did when we discussed the
mechanism of generation of density perturbations. This representation is
only approximate since $D(Y)$ changes a few times when $Y$ changes from
$0$ to $R$. However, as we have already seen, this simple approximation is
very useful if one wants to understand the most important features of the
theory. Let us use the same parameters as in \cite{KOST}. In this case
$D(0) = 100$ and $D(R) = 350$. For definiteness, let us take $D \sim 300$,
which corresponds to the beginning of the process. Then one finds $\phi(x)
\approx 5 M_5^2  Y(x) = 5\cdot 10^{-2} M_p M_5 Y(x)$. According to
(\ref{efflag}), the effective Lagrangian of a properly normalized  field
$\phi$ is given by
\begin{eqnarray}
 L_{\rm eff }&\approx & \frac{1}{2} (\partial_{\mu}\phi)^2  - {\cal V}(\phi)\nonumber\\
  &=& \frac{1}{2} (\partial_{\mu}\phi)^2  + 10^{-22}
M_p^4 \exp \left(-{5000\, \phi\over M_p}\right) \ . \label{propV}
\end{eqnarray}
Exponents such as $\exp {C\phi\over M_p}$ with $C = O(1)$ often appear in
string theory. However,  it is an challenge to find  a realistic model
with a potential  ${\cal V}(\phi)  \propto - 10^{-22} M_p^4 \exp
\left(-{5000\, \phi\over M_p}\right)$. Note that the huge coefficient
$5000$ in the exponent is crucially important for obtaining
long-wavelength perturbations with a nearly flat spectrum in the ekpyrotic
scenario, as well as in the pyrotechnic scenario, see Appendix B.

It is argued in \cite{KOST} that one may think of $V(Y)$ as the potential
derived from the superpotential $W \sim e^{-cY}$ for the modulus $Y$ in
the 4d low energy theory, where $c$ is a positive parameter with dimension
of mass. The corresponding potential in terms of the field $\phi$ is
constructed from $W$ and the K\"ahler potential $K$,
\begin{equation}
V=e^{K/M_{pl}^{2}}\left[K^{ij}D_i W \,  \bar{D_j W}-\frac{3}{M_{pl}^{2}}
W\bar{W}\right], \label{eq:V}
\end{equation}
where $D_i = \partial/\partial \phi^i + K_i/M_{pl}^{2}$, $K_i = \partial
K/ \partial \phi^i$, $K_{ij} = \partial^2 K/ \partial\phi^i
\partial\phi^j$.

Let us indeed try to calculate the corresponding potential. Note that the
properly normalized field $\phi$ is very small, $\phi < 0.05 M_p$ for $Y <
R = M_5^{-1}$. In such a situation one may expect that in the first
approximation $e^{K/M_{pl}^{2}} \approx 1$, $D_i \approx \partial/\partial
\phi^i$. This is what happens if one has minimal K\"ahler potential for
the properly normalized field $\phi$, as suggested by (\ref{propV}). In
reality, the K\"ahler potential may be quite complicated, involving many
other fields, and the superpotential will contain contributions of other
fields in addition to $\phi$. Still it is quite instructive to see what
happens if one considers a single field $\phi$ with the minimal K\"ahler
potential.

To obtain ${\cal V(\phi)} \sim \exp\left(-{5000\, \phi\over M_p}\right)$
one should take  $W(\phi) \sim \exp \left(-{2500\, \phi\over M_p}\right)$.
Then the first, positive term in  (\ref{eq:V}) will be $10^6$ times
greater than the second, negative term, so that the second term can be
neglected, just as  in a globally supersymmetric theory. Therefore, even
though the potential will be proportional to $\exp \left(-{5000\,
\phi\over M_p}\right)$, as  expected in \cite{KOST}, it will be {\it
positive} rather than negative. In this case  the bulk brane will be
attracted to the hidden brane and it will never fall to the visible brane.

The positive sign appears in this expression not by accident. If one would
take $W \sim e^{-c\phi/M_p}$ with   $c \ll 1$, the sign would be negative,
as required. But in this case the scenario would not work because the
spectrum of perturbations would be strongly red. That is why in
\cite{KOST} one has $c \sim 2500$. But in this case the scenario does not
work anyway because the potential becomes positive and the bulk brane does
not move towards the visible brane. It might be possible to resolve this
problem by taking a completely different set of parameters as compared to
the ones taken in \cite{KOST}. Finding a proper set of parameters is a
separate problem to be addressed.

Now let us forget for a while about this problem and simply {\it assume}
that the potential energy of interactions between the  branes produces the
negative potential $V(Y) = -v e^{-\alpha m Y}$, where $Y$ is the distance
between the branes. Indeed, we will see shortly that similar (though
somewhat different)  terms may appear if one considers a contribution of
other matter fields. But then one should  take into account interactions
between the bulk brane and {\it both of the other branes}.  This would add
at least one new term to the potential:
\begin{equation}\label{2pot}
V(Y) \sim -v_1  e^{-c_1 Y} - v_2 e^{-c_2 (R-Y)} \ .
\end{equation}
Here   $c_i$ are some positive constants. The first term describes the
Yukawa-type interaction of the bulk brane with the visible brane, the
second term, which was not present in \cite{KOST}, describes a similar
interaction of the bulk brane with the hidden brane.   Potentials of this
type may indeed appear in a three-brane configuration
\cite{Moore:2000fs}. In general, they may contain many other terms,  the
coefficients $v_i$ may be functions of various moduli, and may be either
positive or negative. Before discussing this more complicated situation
outlined in \cite{Moore:2000fs}, we will discuss our toy potential
(\ref{2pot}) to develop some intuition.

The appearance of the second term in the expression for $V(Y)$ is very
important. Now  the potential near the hidden brane is entirely dominated
not by the exponentially small term $-v_1 \exp (-c_1 Y)$, but by the
second term $- v_2 \exp (-c_2 (R-Y))$. If both $v_1$ and $v_2$ are
positive, then the bulk brane is attracted to the hidden brane and  never
moves towards our brane. Meanwhile, if $v_2$ is negative, there will be a
large repulsive force  between the hidden brane and the bulk brane. As a
result, the bulk brane will be rapidly moving towards the visible brane.
The total duration of the process will be very short, and therefore no
long wavelength perturbations $\delta Y(k)$ will be produced.

The only way to overcome this problem would be to have the second
exponential term extremely small. In this case the bulk brane could stay
for a while in the shallow minimum of $V(Y)$ near $Y = R$, then tunnel
through the barrier and fall to our brane. However, it must tunnel to the

part of the potential with the curvature smaller than $k_0^2 \sim 10^{-64}
M_5^2$ if one wants to produce density perturbations on the scale of the
present horizon. Simple estimates indicate that it would only be possible
if $v_2 <  10^{-60}$. In other words, one should completely forbid any
contribution to the superpotential coming from the interaction of the bulk
brane with the hidden brane.

An example of the calculation of the effective potential $V(Y)$  due to
the nonperturbative instanton effects in HW theory was given in
\cite{Moore:2000fs}.   The  calculation was very complicated, and it was
based on several carefully specified assumptions.  In particular, they
assumed that the contributions of  the hidden and of visible branes to the
superpotential are of comparable magnitude, and they  included
contributions of just a  few matter fields assuming  special relations
between their values. Still their results are very interesting and
instructive. We will present them in notation of \cite{Moore:2000fs}. The
nonperturbative potential is schematically represented as
\begin{eqnarray}\label{schempot}
 U &=&
\frac{1}{V J^2} \Biggl(\alpha C^4 - \beta(1-x) \vert C\vert^3 \vert
 e^{-J x} \mp
 e^{-J(1-x)} \vert  \nonumber\\
 &+& \gamma V \Bigl [ \bigl(e^{-Jx} \mp e^{-J(1-x)}\bigr )^2+
\frac{2J}{3V}\bigl (1-2x\bigr )e^{-2J(1-x)}
 \pm \frac{4Jx}{3V}e^{-J} \Bigr ]\Biggr).
\end{eqnarray}
Here $V$ and $J$ are some slowly moving moduli, $C$ are charged scalars
living on the visible brane, $x$ is the bulk brane coordinate changing
from $0$ (visible brane) to $1$ (hidden brane). This result is valid under
several conditions including the requirement $Jx \gg 1$, $J(1-x) \gg 1$
(i.e. the exponents $e^{-Jx}$ and $e^{-J(1-x)}$ must be exponentially
small indeed). The exponent $e^{-Jx}$ in notation of \cite{Moore:2000fs}
corresponds to $e^{-\alpha m Y}$ in \cite{KOST}.

This expression has  positive and negative terms, with exponential and
non-exponential factors which may rise or fall near each of the branes.
The   term $\alpha C^4$ is positive; it does not depend on $x$ (i.e. on
$Y$ in notation of \cite{KOST}). The term $- \beta(1-x) \vert C\vert^3
 e^{-J x}$, which appears  due to interference of the nonperturbative superpotential with the superpotential of charged scalars $C$, is the only  term with the desired behavior $\sim -e^{-J x}$. However, it is shown in \cite{Moore:2000fs} that this term is subdominant and  {\it the sum of all terms is always  positive}  within the domain of validity of Eq. (\ref{schempot}).  To obtain the negative exponential potential required in the ekpyrotic scenario one would need to forbid all  terms except the negative term $\sim -e^{-J x}$ in Eq. (\ref{schempot}). This term must  appear because of the interaction of the bulk brane with the visible brane where the group $E_8$ is broken. In particular, we would need to forbid all terms $\sim \pm e^{-J(1-x)}$ that would appear because of the interaction of the bulk brane with the $E_8$ brane.

This is a rather nontrivial task. According to \cite{LOP}, the
nonperturbative contribution to the superpotential is only nonvanishing if
the restriction of  vector bundle to the holomorphic curve around which
the supermembrane is wrapped is trivial. This condition is satisfied when
the bulk brane interacts with  the end-of-the-world (hidden) brane with
unbroken $E_8$.  That is why the superpotential calculated in \cite{LOP}
described interaction of the bulk brane with the $E_8$ hidden brane, but
not with the visible brane. But in our case such interaction would induce
the term  $\sim e^{-c_2 (R-Y)}$, which should be forbidden  in the
ekpyrotic scenario. Does it mean that this scenario requires that our
brane was in fact the $E_8$ brane before the brane collision, and then
$E_8$  was broken down to the symmetry of the standard model after the
collision of the brane with tension $\alpha$ with the brane with an
extremely small tension $\beta = \alpha/2500$? So far no realization of
such a scenario was proposed. All previous works on this subject assumed
that we live on the  brane where $E_8$ was already broken to some smaller
group prior to the collision, and that the colliding branes had comparable
tensions, which is not the case in \cite{KOST}.

Now let us look at this situation from a somewhat different perspective.
Historically, one of the main reasons to calculate the nonperturbative
brane potential was to find a mechanism of brane stabilization in the HW
scenario. Indeed, at the classical level these branes can stay at any
distance from each other, as long as no naked  singularity appears between
the branes. The hope was that the distance between the branes will be
stabilized due to nonperturbative effects. The result of the calculations
performed in \cite{Moore:2000fs} shows that under the conditions specified
in this work the nonperturbative effects instead of the brane
stabilization produce a small destabilizing {\it repulsion} between the
branes, proportional to $e^{-J}$. In the language of the ekpyrotic
variables, this would correspond to the repulsion proportional to $
e^{-\alpha m R} \sim e^{-250}$.

This result has several important implications. First of all, at present
the problem of brane stabilization in the HW scenario remains unsolved.
Second, if the brane stabilization occurs due to the nonperturbative
effects considered in \cite{Moore:2000fs}, then the stabilizing forces
will be vanishingly small if one uses the parameters of the ekpyrotic
scenario. There is no much freedom in making $ e^{-\alpha m R}$ larger
because the absolute value of the curvature of the effective potential
${\cal V}(\phi)$ near the hidden brane  must be smaller than $k_0^2 \sim
10^{-64} M_5^2$ if one wants to produce density perturbations by the
mechanism of tachyonic instability, see Appendix B. Thus we really need to
have $\alpha m R >  10^2$. Consequently, the $T$ moduli (or the $\phi$
field in our notation) corresponding to the brane excitations will be
nearly massless. This does not seem to be phenomenologically acceptable.
In addition, in the absence of a sufficiently powerful mechanism of brane
stabilization  there is no obvious reason to expect that the  branes must
be parallel to each other from the very beginning. In the beginning of the
evolution of the universe different parts of the branes at large distance
from each other ``did not know'' where they should stay. As we will see in
the next section, this leads to a severe problem of homogeneity.

On the other hand, if eventually we will discover the mechanism of brane
stabilization in the HW scenario, then most probably this mechanism will
apply not only to the visible and hidden branes, but to the bulk brane as
well. This will imply that the potential ${\cal V}(\phi)$ will acquire
much greater curvature than $10^{-64} M_5^2$, in which case the mechanism
of tachyonic instability will be unable to generate large-scale density
perturbations.

Can we have the best of both worlds: brane stabilization and large-scale
density perturbations? Yes, we hope that it is possible, but only if we
have inflation.

\section{ Homogeneity and entropy problems}\label{hom}

Now let us be very optimistic and assume that all previously mentioned
problems can be resolved  and let us see whether this scenario can solve
the homogeneity problem. Of course, one may {\it assume} that the universe
from the very beginning was entirely homogeneous. The idea is that our
universe starts its evolution in a BPS state, which is a completely stable
lowest energy state containing two homogeneous branes.

Here we have several important issues at once. First of all, one may
indeed expect that the universe after a long and violent evolution {\it
ends up} in a ground state. Thus, one could argue that the {\it final}
state of the universe, rather than its {\it initial} state, could be a BPS
state. But is it possible to start with the universe being in a ground
state? Is it possible that a ground state of a theory decays? A decaying
state cannot be a true ground state. The existence of the non-vanishing potential $V(Y)$ violates the BPS nature of the initial state and leads to the instability the brane configuration which develops within {\it finite time}.  But  a  nonsingular  state with finite lifetime cannot be a true initial state of the universe, at least not in the classical  theory of gravity.

Essentially we have two different options. The first one is that the
universe appeared from an initial singularity or was created ``from
nothing,'' then it experienced a period of expansion, cooling down, and
eventually reached its ground state. Here, there is no obvious reason to
expect that it begins in a ground state.

Another option is that we are in a self-reproducing false vacuum state.
There is no initial singularity, and there is an ongoing repetitive
process of creation of new parts of the universe. The first semi-realistic
version of this idea was proposed in \cite{Gott,Et82} in the context of
the eternal new inflation scenario. However, it was immediately realized
that this idea will not work and the universe in the new inflation
scenario must have a beginning because of geodesic incompleteness of an
expanding de Sitter universe \cite{Linde:1982gg}, see also
\cite{VilBorde}. A similar conclusion may not be valid for chaotic
inflation \cite{NoTheorem}, so it remains to be seen whether eternal
inflation in the simplest versions of the chaotic inflation scenario
\cite{Eternal} requires any beginning.

However, in the ekpyrotic scenario there is no inflation, by design, and
no self-reproduction. The properties of the BPS state (the tension of the
branes) change each time a new bulk brane is born. Thus, it is not a
stationary process, so one cannot avoid the question of the initial
conditions that could create the two- or three-brane near-BPS state. In
such a case, the required homogeneity of the two- or three-brane universe
should be explained rather than postulated. The initial homogeneity of the
two- or three-brane configuration postulated in the ekpyrotic scenario is
not a solution of the problem but rather a problem to be solved.

Of course, it may happen that an approximate homogeneity on a very small
scale is all  we need. This is the case in the simplest versions of
chaotic inflation where an approximate homogeneity on the Planck scale is
the only condition required to trigger an eternal chain reaction of
self-reproduction of the inflationary universe \cite{Eternal}.  Let us see
whether an approximate homogeneity on small scales is good enough for the
consistency of the ekpyrotic scenario.

Suppose that initially the position of the bulk brane  was slightly
perturbed, so that it was equal to $Y(0) + \Delta Y(x)$, with $Y(0)
\approx R$. For simplicity one may assume that this perturbation can be
represented by a sinusoidal wave $\Delta Y(k) \sin kx$. Then  we
immediately see a possible problem: A very small {\it classical}
inhomogeneity $\Delta Y(x)$ of the initial position of the bulk brane can
be exponentially enhanced by the tachyonic instability, which may lead to
a strong inhomogeneity of the visible brane upon collision. In other
words, {\it the same mechanism that produces large scale classical
perturbations from small quantum fluctuations may greatly amplify small
initial inhomogeneities of the position of the bulk brane}. If the
amplitude of the {\it classical} perturbations $\Delta Y(k)$ is greater
than the average amplitude of quantum fluctuations $\delta Y(k)$, we will
see unacceptably large irregularities in the CMB spectrum. To avoid this
problem we must require that the classical perturbations of the bulk brane
position $\Delta Y(k)$ are smaller than the quantum perturbations $\delta
Y(k)$ for all wavelengths that we can presently observe.\footnote{One may
even argue that the requirement that classical perturbations must be
smaller than the quantum ones means that strictly speaking there are no
classical perturbations at all. Indeed, perturbations can be called
classical only if the corresponding occupation numbers $n_k$ for particles
with momenta $k$ are much greater than $1$. But then the amplitude of such
perturbations would become greater than the amplitude of quantum
fluctuations by a factor of $\sqrt {2n_k +1}$ \cite{book}, which is
incompatible with the condition  $\Delta Y(k) < \delta Y(k)$. In this
sense one may say that to avoid large CMB anisotropy one should not have
{\it any} large-scale classical perturbations of the bulk brane: We must
start with an ideally flat brane.}

To evaluate the significance of this effect we will estimate the initial
amplitude of  quantum fluctuations $\delta Y(k)$ on the scale
corresponding to our present horizon, $l_0 \sim 10^{28}$ cm. As we have
shown in the previous section, at the moment of the brane collision such
perturbations had momentum $k_0 \sim 10^{-17}$ GeV $\sim 10^{-32} M_5$.
The corresponding length scale $k_0^{-1}$ was $10^{30}$ times larger than
the proper distance between the branes $R \sim B D^2 M_5^{-1} \sim 10^2
M_5^{-1}$.

Eq. (\ref{Yk}) gives the following expression for the average amplitude of
quantum fluctuations $\delta Y(k_0)$:
\begin{equation}
\delta Y(k_0) \sim   {k_0  \over   \sqrt{ \beta M_5^3 B} D }   \sim
10^{-32} M_5^{-1} .
\end{equation}
To avoid problems with anomalously large  CMB anisotropy one should have
$\Delta Y(k_0) < \delta Y(k_0)\sim 10^{-32} M_5^{-1}$. Dividing this by
the initial value $Y\approx R = M_5^{-1}$ one finds that in order to avoid
unacceptably large density perturbations and CMB anisotropy in the
observable part of the universe one must have the branes positioned
exactly parallel to each other with accuracy
 \begin{equation}\label{inhom}
{\Delta Y(k_0)  \over Y}  \sim 10^{-32}
\end{equation}
on the macroscopically large scale $k_0^{-1} \sim 10^4$ cm, which is 30
orders of magnitude greater than the physical distance between the branes
$B D^2 M_5^{-1}$.\footnote{The result  (\ref{inhom}) can also  be  derived
using variation with respect to time in formula (\ref{time2}).} In other
words, the angle $\theta$ between the branes on the scale $k_0^{-1} \sim
10^{32} M_5^{-1}$  must be smaller than $10^{-62}$ from the very
beginning.
 This incredible fine-tuning cannot be considered a solution of the homogeneity problem.

Can we do something about it? One possible idea would be to deviate from
the static setting describing initial configuration of two or three branes
in a near BPS state, as suggested in \cite{KOST}, and instead consider the
process of cosmological evolution which could eventually result in
creation of such a configuration. For example, one may imagine that
initially there was a stage of inflation which made the universe
homogeneous and the branes parallel.  Another possibility is to consider a
non-inflationary Friedmann  evolution starting with a cosmic singularity
and resulting in formation of two branes. If there exists a powerful
mechanism of brane stabilization, then the branes  could stay at a
distance approximately equal to $M_5^{-1}$  for an exponentially long
time. Then the energy density of matter on the branes, including the
energy of their inhomogeneities, will be diluted by cosmic expansion, and
eventually the branes will become almost exactly parallel to each other.

Let us assume for a moment that we were able to make the universe
homogeneous by this mechanism (which is not a part of the ekpyrotic
scenario assuming static initial conditions). Could we solve all major
cosmological problems due to the stage of the cosmological expansion
preceding the onset of the pyrotechnic stage? Suppose that the universe is
closed, and initially it was filled with radiation. Then, according to
\cite{book}, its total lifetime is given by $t \sim S^{2/3} M_p^{-1}$,
after which it collapses. Ignoring for a moment the possible
time-dependence of $M_p$, we find that in order to survive until the
moment $t \sim k_0 \sim 10^{32} M_5 \sim 10^{34} M_p$, the universe must
have the total entropy greater than $10^{50}$. In other words, the
universe must contain at least $10^{50}$ elementary particles from the
very beginning. Thus in order to explain why the total entropy (or the
total number of particles) in the observable part of the universe is
greater than $10^{88}$ one must assume that it was greater than $10^{50}$
from the very beginning. This is the so-called entropy problem
\cite{book}. If the universe initially has the Planckian temperature, its
total initial mass must be greater than $10^{50} M_p$.

On the other hand, if the bulk brane  was created due to the tunneling,
then one must make sure that the tunneling is extremely strongly
suppressed so that it does not happen during an exponentially large time
required for the branes to become parallel with accuracy of $10^{-62}$.
One must also ensure that the tunneling does not take place twice within
the time $10^{32} M_5^{-1}$ that is required for the formation of our part
of the universe. Indeed, each tunneling makes the universe inhomogeneous,
but if only one such event occurred, it may be interpreted as a creation
of a homogeneous open universe. However, such interpretation will fail if
there were many bubbles within the cosmological horizon.

More importantly,  our mechanism of brane ``homogenization'' could work
only if there were some reason for the branes to be dynamically stabilized
immediately after the beginning of the evolution of the universe, at the
same distance all over the huge domain  many orders of magnitude greater
than the brane separation. However, the problem of brane stabilization in
the HW scenario still remains unsolved, and in the ekpyrotic scenario
there are no forces that would keep the branes at a fixed distance. It is
possible that nonperturbative effects similar to those responsible for
generation of the potential $V(Y)$ will fix the distance between the
branes \cite{HWph,Moore:2000fs,Banks:1999ay}. But with the parameters used
in \cite{KOST} such stabilizing forces would be suppressed by the same
kind of exponents as $V(R) \sim 10^{-120}$, i.e. they will be incredibly
weak.

Meanwhile, if the branes were even slightly inhomogeneous from the very
beginning,  they were out of the BPS regime, and the  long-range forces of
attraction and repulsion were not compensated \cite{Ts,DvSh}.   Our
estimates indicate that if the initial angle between the branes was
greater than  $10^{-62}$, these forces could be much stronger than the
nonperturbative potential $V(Y)$. The potential $V(Y)$ would contain  a
large power-law contribution \cite{DvaliTye} which would lead to a
premature fall of the bulk brane to our brane. In such a case the
perturbations with a flat spectrum would not be produced.

Moreover, if one considers a generic inhomogeneous regime in the early
universe, where the initial fluctuations of metric could be $O(1)$ on the
Planckian scale, and the branes were not parallel at all, then  the
non-BPS long range forces of attraction and repulsion could be dozens of
orders of magnitude greater than  $V(R)\sim 10^{-120}$. In this case we do
not see any way to make the universe even marginally homogeneous on the
scale $10^{30}$ times greater than the brane separation.

In comparison, in the simplest versions of chaotic inflation scenario the
homogeneity problem is solved if our part of the universe initially was
relatively homogeneous on the smallest possible scale $O(M_p^{-1})$
\cite{chaot}. The whole universe could have originated from a  domain with
total entropy $O(1)$ and  total mass $O(M_p)$. Once this process begins,
it leads to eternal self-reproduction of the universe in all its possible
forms \cite{Eternal,book}. Nothing like that is possible in the ekpyrotic
scenario.

\section{Conclusions}

In this paper we were trying to evaluate the claims that the recently
proposed ekpyrotic scenario is fully motivated by string theory and
resolves all major cosmological problems without using inflation. These
are very serious claims since so far all attempts to replace inflation by
an equally valid paradigm were unsuccessful.

In our opinion, this new attempt is not an exception.  We have found that
the ekpyrotic scenario is not entirely string motivated and must be
modified. In particular, to obtain this scenario from the
Ho\u{r}ava-Witten model one must change the signs of the brane tensions,
which entails many other changes in the parameters and properties of the
model.

One must also find  the way to generate the brane potential $V(Y)$, which
in terms of the effective field $\phi$ is given by a very unusual
expression $-10^{-22} M_p^4 \exp \left(-{5000 \phi\over M_p}\right)$. To
find such potentials one must allow nonperturbative interaction of the
bulk brane with the visible brane and entirely suppress interaction of the
bulk brane with the hidden brane. This is a requirement which is difficult
to satisfy. And in the end one would need  this potential to vanish at
$\phi = 0$. As we have argued, existence of such potentials is hardly
compatible with string phenomenology and with the possibility to achieve
brane stabilization in the HW scenario.

Many other features of this model are equally questionable. Is it really
possible for BPS states to decay?  Does the bulk brane have flat geometry?
What exactly happens when the branes collide? If the bulk brane brings too much
non-expanding matter to our world, our universe may collapse rather than
expand. Indeed, prior to the collision, our brane  was empty. If one simply deploys there a lot of matter, the universe may collapse. It is not sufficient to reheat our brane and create matter there. One must make sure that this
matter expands rather than implodes, and that it expands in such a way
that the kinetic energy of matter is exactly equal to its potential
energy, because otherwise our universe will not be flat, and there will be no inflation to make it flat later.

To study this  problem one would need to use the Israel junction
conditions for the extrinsic curvature $K^{\mu}_{\nu}$ for the colliding
branes embedded into 5d bulk. Usually the hypersurface $\Sigma$ in the
junction conditions is a time-like hypersurface. In this model we have to
consider a space-like hypersurface $t_0$, where the four-dimensional
brane energy momentum tensor $T^{\mu}_{\nu}$ experiences a jump from the
vacuum-like form $T^{\mu}_{\nu}=\sigma \delta^{\mu}_{\nu}  $ to the
radiation form $T^{\mu}_{\nu}=\rho\ {\rm diag}( 1, -1/3,-1/3,-1/3)$.
Therefore the  extrinsic curvature (embedding) of the visible brane also
will have a jump at $t_0$. This means that cosmology at the visible brane
after the collision may be more complicated than what one may naively
expect.\footnote{Recently it was found that the 5d description of the ekpyrotic scenario  is problematic even before the collision  \cite{KKLT}. It was shown there that the ansatz for the metric and the fields used in \cite{KOST} does not provide a consistent solution to the dilaton and gravitational equations in the bulk. To avoid this problem one would need to use a more general ansatz for the metric  and  provide an improved 5d interpretation of the bulk brane potential $V(Y)$.}

All of these issues are very non-trivial. Our experience with brane
cosmology tells us that it is often dangerous to make approximations which
at the first glance seem very natural. For example, we have found that if
one takes the two-brane Randall-Sundrum model and adds there a scalar
field in order to achieve brane stabilization without changing the brane
tension, as in the Goldberger-Wise scenario \cite{GW}, the branes become
exponentially expanding \cite{GKL}. To avoid this expansion one must
adjust the brane tensions in a specific fine-tuned way \cite{DFGK}. We
suspect that a similar adjustment is necessary in the
ekpyrotic/pyrotechnic scenario as well.

 But the main problem is  related to the  claims that the ekpyrotic scenario provides us with a
new non-inflationary brane-specific mechanism of generation of density
perturbations with nearly flat spectrum, and that it also provides a
solution to the homogeneity, horizon and flatness problems. In this paper
we have shown that there is nothing brane-specific in the ekpyrotic
mechanism of production of density perturbation. It is based on the simple
mechanism of tachyonic instability, which works in 4d theory as well. But
to make it realistic one must consider a narrow subclass of potentials
with $V'' <0$ that would lead to inflation if their maxima would
correspond to $V > 0$. Then if one wants to avoid inflation  one must
fine-tune the height of the maximum  with accuracy about $10^{-120}$.
Finally, one must ensure that with this setting we do not wind up in AdS
universe with large negative cosmological constant.

It is ironic that if this goal is achieved, we will unleash a mechanism of
tachyonic instability which will exponentially amplify not only quantum
fluctuations, but also initial inhomogeneities.

To understand the nature of  the problem  one may compare this scenario
with inflation. Consider, for example, a potential with $V''(\phi) < 0$
used in new inflation. Inflation occurs if $|V''| \ll H^2$. Therefore the
exponential tachyonic instability, which is controlled by $\sqrt {|V''|}$
(and dampened by inflation) develops much more slowly  than the
exponential stretching of the universe controlled by $H$: $\delta\phi \sim
\exp \left({|V''|\over 3 H}t\right)$, whereas $a \sim \exp (Ht)$. As a
result, all perturbations which could exist prior to inflation are
stretched away. This combination of two instabilities dominated by
expansion is a unique and very important property of inflation.

Meanwhile in the ekpyrotic scenario the only instability is the tachyonic
one. If it is powerful enough to produce classical perturbations out of
quantum fluctuations, it is equally powerful in making  small classical
perturbations exponentially large. To see CMB anisotropy generated by
quantum fluctuations but not by initial inhomogeneities, the initial
inhomogeneities must be below the level of quantum noise.

Thus, inflation removes all previously existing inhomogeneities, whereas
in the ekpyrotic scenario even very small initial inhomogeneities become
exponentially large. Therefore instead of resolving the homogeneity
problem, it makes this problem much worse. Moreover, if the universe in
this scenario was even slightly inhomogeneous, the long-range forces,
which would be cancelled in a BPS state, appear again. This may completely
change  the whole scenario.

To avoid this problem one must provide a physical mechanism which would
make the branes  parallel to each other with accuracy $10^{-60} $ on the
scale   30 orders of magnitude greater than the distance between the
branes.

In addition, in order to survive until the horizon  becomes 30 orders of
magnitude greater than the distance between the branes, our universe from
the very beginning must have entropy greater than $10^{50}$, which
constitutes the entropy problem.

This demonstrates once again how difficult it is to construct a consistent
cosmological theory without using inflation.

\

The authors are grateful to T. Banks, M. Dine, P. Greene, S. Dimopoulos,
G. Dvali, N. Kaloper, G. Moore, P. Nilles, A. Peet, G. Peradze,
N.~Saulina, G. Shiu and L. Susskind  for important comments. We thank NATO
Linkage Grant 975389 for support.  L.K. was supported by NSERC and by
CIAR; R.K. and  A.L. and  were supported by NSF grant PHY-9870115; A.L.
was also supported  by the Templeton Foundation.

\

{\bf Note Added:} Recently the authors of the ekpyrotic scenario issued a paper replying to some of our comments on their theory \cite{Khoury:2001iy}. However, since they incorporated many of our results and suggestions in the revised version of their paper  \cite{KOST}, we no not think that there is any real disagreement with respect to our results.

The only ``incorrect'' statement  they have found in our work was our conclusion that  the HW phenomenology requires visible brane with positive tension. However,  it is definitely true that this requirement is satisfied in all works on the HW phenomenology \cite{Witten,HWph,lukas} to which the authors of \cite{KOST} referred in their paper.  The only exception that we are aware of is provided by the unconventional approach to the HW phenomenology outlined in \cite{Benakli:1999sy,Donagi:2001fs}; see a detailed discussion of this issue in Section \ref{super} of our paper and in \cite{KKLT}. Instead of repeating this discussion here, we just mention that the authors of \cite{KOST} removed the statement that the visible brane must be in the small-volume region of space-time (i.e. that it must have negative tension) from the revised version of their paper \cite{KOST}. They also removed  the ``justification'' of this statement in Section VB.  After that they said  \cite{Khoury:2001iy} that they never claimed that the tension of the visible brane must be negative.

Another point of criticism was related to our use of the theory of tachyonic preheating \cite{tach} for the derivation of the amplitude of the density perturbations in the ekpyrotic scenario. This derivation  allowed us to show that the assumption that $D(Y)$ must decrease towards the visible brane was not necessary for generation of density perturbations. This assumption was the basis for the statement that the density perturbations in the ekpyrotic scenario have blue spectrum \cite{KOST}. We  also found an error by a factor of $(3B)^{-1/2} \sim 20$ in the expression for density perturbations in Eq. (75) of \cite{KOST}. After that, the authors of \cite{KOST} have withdrawn the statement that the spectrum of the density perturbations in the ekpyrotic scenario must be blue. They improved Eq. (75),  and made a dramatic modification of all parameters of their model in order to keep it consistent with the observational data. 

 The remaining points of disagreement  are rather philosophical. For example, it is argued in \cite{Khoury:2001iy} that until a theory of quantum gravity is fully developed, we will not know which initial conditions are better. However, we still believe that it is much easier to imagine that the early universe was relatively homogeneous on the  Planck scale $O(10^{-33})$ cm, as required for the eternal process of chaotic inflation to begin \cite{book}, than to assume that the universe from the very beginning was huge and nearly exactly homogeneous on a scale $10^{30}$ times greater than the Planck scale.

\

\section*{Appendix A. Spectrum of fluctuations produced by the tachyonic instability}

Let us derive more rigorously the spectrum of fluctuations considered in
Section IV. We shall
 consider the equation for quantum fluctuations (\ref{tach}).
Let us begin with a class of power-law potentials $V(\phi)=-\lambda
\phi^n/M^{n-4}$, $n >2$. The equation for
 the temporal part of the eigenmode function
 $ \delta\phi_k(t) e^{-i{\vec k \vec  x}}$
is
\begin{equation}\label{tach1}
 \ddot{\delta\phi_k}+\left( k^2
 -n(n-1) \lambda {\phi^{n-2} \over M^{n-4}}  \right)\delta\phi_k =0  \ .
\end{equation}
Initially the field rolls from the top of the potential, $\phi \approx 0$,
towards large $\phi$, $\dot \phi >0$ (in the context of the moduli field
between branes we shall simply redefine $\phi \to \phi_0-\phi$). Assuming
that the initial energy of background field  vanishes, we find the time
evolution of the background field in the form
\begin{equation}\label{time1}
t_0-t= { 2 \over {n-2}} { M^{(n-4)/2} \over \sqrt{2\lambda
}}\left(\phi(t)^{(2-n)/2}- \phi_0^{(2-n)/2} \right).
\end{equation}
We will consider the evolution of the field until it hits a certain value
$\phi_0$. It is convenient to choose $t_0=  { 2 \over {n-2}}{ M^{(n-4)/2}
\over \sqrt{2\lambda }}\phi_0^{(2-n)/2}$, then time $t$ flows from the
initial value to $t_0$. Substituting  expression (\ref{time1})  into eq.
(\ref{tach1}), we reduce (\ref{tach1}) to the Bessel equation
\begin{equation}\label{hankel1}
 \ddot{\delta\phi_k}+\left( k^2  -{ {n(n-1)} \over (n-2)^2} \cdot { 2 \over
{t^2}}\right)\delta\phi_k =0 \ .
\end{equation}
A solution  corresponding to the positive-frequency initial vacuum
fluctuations in given in terms of the Hankel function
\begin{equation}\label{solution1}
\delta\phi_k(t)=N \sqrt{t} { \cal H}^{(2)}_\mu ( k t  )
\end{equation}
 with the index
\begin{equation}\label{mu}
\mu^2=1/4+2{ {n(n-1)} \over (n-2)^2} .
 \end{equation}
Early time asymptotic value for the large argument $kt$ should be
$\delta\phi_k(t)={ 1 \over \sqrt{2k}} e^{-ikt}$, so we  choose the
normalization factor $N={\sqrt{\pi} \over 2}e^{-i\pi \mu/2}$. The spectrum
of fluctuations at the  moment $t_0$ is given by the expression
(\ref{solution1}) at $t_0$ as a function of $k$. For large $kt_0  >1$ the
spectrum will be an oscillating function of $k$. However, for large
wavelengths fluctuations with $k< t_0={ 2 \over {n-2}}{ M^{(n-4)/2} \over
\sqrt{2\lambda }}\phi_0^{(2-n)/2}$, the fluctuations are frozen with the
amplitude  which can be estimated from the small argument asymptotic of
the Hankel function,
\begin{equation}\label{asymp1}
\vert \delta\phi_k(t_0)\vert \sim k^{-\mu} \ .
\end{equation}
Further, we can find the spectrum of density fluctuations in the large
wavelength limit. The density fluctuations are given by $\delta_k \simeq H
{{\delta \phi_k(t_0)} \over {\dot \phi(t_0)}} ({k \over {2\pi}})^{3/2}$.
The only $k$ dependence is in the $\delta \phi_k(t_0)$. Therefore the
spectrum of the density fluctuations $\vert \delta_k \vert^2 \sim
k^{3-2\mu}$. We will use the spectral index of the density fluctuations,
which is defined by the formula $\vert \delta_k \vert^2 \sim k^{(n_s-1)}$.
We have
\begin{equation}\label{index1}
n_s=1+(3-2\mu) \ .
\end{equation}
Substituting here expression (\ref{mu}) for $\mu$, we find  simple formula
\begin{equation}
n_s=1-{4 \over {(n-2)}} .
\end{equation}
Remarkably, this is precisely the same result as derived in Section IV by
elementary methods. We see that the spectrum in the class of the power-law
potentials  is always red, $n_s <1$.

In the limit $n \to \infty$, which can be considered as a shortcut answer
for the exponential potential, one may expect the flat spectrum, $n_s=1$.
Indeed, let us consider this {\it very specific} potential $V=-V_0
e^{-\phi/M}$, which plays a special role in the ekpyrotic scenario. We
shall consider the evolution of quantum fluctuations at the time when the
field $\phi$  rolls from the top of the potential  (i.e. from large
$\phi$) to $\phi=0$. For the temporal part of the eigenmode function
 $ \delta\phi_k(t)$ we have
\begin{equation}\label{tach2}
 \ddot{\delta\phi_k}+\left( k^2
 -{ V_0  \over M^2} e^{-\phi/M} \right)\delta\phi_k =0  \ .
\end{equation}
Assuming that the initial energy of background field is vanishingly small,
we find
\begin{equation}\label{time2}
-t+t_0={ M \over \sqrt{V_0/2}}\left(e^{\phi/2M} -1\right).
\end{equation}
Here $\phi(t_0)=0$, and the time $t$  flows from $-\infty$  to $t_0$ as
the field $\phi$ rolls from the top of the potential  (large $\phi$) to
$\phi=0$. It is convenient to choose $t_0=- { M \over \sqrt{V_0/2}}$, then
$ e^{-\phi/M}={ 2M^2 \over V_0}{ 1 \over t^2}$. Substituting this
expression into eq. (\ref{tach2}), we obtain
\begin{equation}\label{hankel2}
 \ddot{\delta\phi_k}+\left( k^2  -{ 2  \over  t^2}\right)\delta\phi_k=0  \ .
\end{equation}
A solution  corresponding to the positive-frequency initial vacuum
fluctuations in given in terms of the Hankel function
\begin{equation}\label{solution2}
\delta\phi_k(t)=N_1 t^{1/2}{ \cal H}^{(1)}_{3/2} (k \vert t \vert ) \ ,
\end{equation}
where ${\cal H}^{(1)}_{3/2}(z)= -\sqrt{2 \over {\pi z}}e^{iz}(1+{i \over
z})$, and we choose the normalization factor $N_1=-{\sqrt{\pi} \over 2}$.
Indeed, for $\vert t \vert \to \infty$ we have $\delta\phi_k(t)={ 1 \over
\sqrt{2k}} e^{ikt}$ (time is negative). However, the most interesting
asymptotic corresponds to the moment $t_0$. For the modes $k \vert t_0
\vert ={{k M} \over \sqrt{V_0/2}} \ll 1$, we have
\begin{eqnarray}\label{asymp2}
&&\vert \delta\phi_k(t_0) \vert={ 1
 \over {\sqrt{2} \vert t_0 \vert  k^{3/2}}}
={\sqrt{V_0} \over {2M}}{ 1 \over k^{3/2}} ,  \nonumber\\
&& \delta\phi(k)=({k \over {2\pi}})^{3/2}\vert \delta\phi_k(t_0) \vert
\simeq {\sqrt{V_0} \over M}  \ .
\end{eqnarray}

Cosmological fluctuations would have a flat spectrum with the amplitude
\begin{equation}\label{cosm}
\delta_k \simeq  {H \over M}  \ .
\end{equation}
The spectrum of density fluctuations will be exactly flat, $n_s=1$.

\section*{Appendix B. The choice of parameters and the bulk brane potential $V(Y)$ in the pyrotechnic scenario}

As we have shown in Section \ref{pot}, the   effective potential  of a
properly normalized  field $\phi$ in the ekpyrotic scenario has a very
peculiar form
\begin{eqnarray}
 {\cal V}(\phi) \sim - 3\beta B M_5^3  v \exp \left(-{\alpha m  \phi\over \sqrt {3\beta B M_5^3}\, D }\right) \sim - 10^{-22}
M_p^4 \exp \left(-{5000\, \phi\over M_p}\right) \ . \label{propV1}
\end{eqnarray}
The huge coefficient $5000$ in the exponent makes this scenario rather
suspicious from the point of view of string phenomenology. Usually the
coefficients which appear in the exponential potentials in string theory
are $O(1)$.

This large number appears when we are using the same parameters as in the
ekpyrotic scenario \cite{KOST}. One may wonder whether it is possible to
obtain a much smaller number by a proper choice of the parameters. Indeed,
one must change these parameters anyway if one wants to describe a
realistic model with $D(Y)$ decreasing towards the hidden brane, as in the
pyrotechnic scenario.

In order to answer this question we will first outline some (though not
all)  requirements for the parameters of the pyrotechnic scenario. Here
$\alpha$ is a {\it positive} tension of the visible brane. In this
scenario one has
\begin{equation}\label{DY}
D(Y) = C - \alpha Y \ .
\end{equation}
Thus one should have
\begin{equation}\label{DY2}
\alpha M_5^{-1} = \alpha R  < C \ .
\end{equation}

The absolute value of the second derivative of the effective potential
${\cal V}(\phi)$, which is given by $\alpha^2 m^2 D^{-2} v e^{- \alpha m
Y}$, determines the square of the comoving momenta $k^2$ of the
fluctuations generated due to the tachyonic instability. This quantity
should be further divided by $D$ to obtain the square of the  physical
momentum \cite{KOST}.  One can obtain density perturbations on the scale
comparable to the size of the observable part of the universe if $\alpha^2
m^2 D^{-3} v e^{- \alpha m R} < k_0^2$, where $k_0$ is the momentum
corresponding to our present horizon, $k_0^2  \sim 10^{-64} M_5^2  \sim
e^{-150} M_5^2$, see Eq. (\ref{k0}). If this condition is not satisfied,
there will be no large-scale structure in the observable part of the
universe. The expression for $k_0$ can be a few orders smaller or greater,
depending on the choice of the parameters, but this will not affect our
final estimate in a noticeable way. Combining all numbers together, we get
the following  constraint on $\alpha m M_5$:
\begin{equation}\label{almm}
\alpha m R = \alpha m M_5^{-1} \gtrsim 120\ .
\end{equation}
Thus the exponent $e^{-\alpha m Y}$ near the hidden brane must be smaller
than $e^{-120}$ if we want to explain the large-scale structure of the
universe by the tachyonic instability. This is an important constraint,
since it shows that all corrections to $V(Y)$ near the hidden brane must
be suppressed with accuracy $e^{-120} \sim 10^{-50}$.

Another important relation that we are going to use is $M_p^2  = BM_5^3
I_3(0)$, where for $\beta \ll \alpha$ one has $I_3(0) = {1\over
2\alpha}[D^4(0)-D^4(R)] $ \cite{KOST}. Since in the realistic theory one
must have $D(R)$ several times smaller than $D(0) = C$, in the first
approximation one has $ I_3(0) \approx {C^4\over 2\alpha}$. Consequently,
\begin{equation}\label{mpl}
M_p  \approx {\sqrt{BM_5^3} C^2\over \sqrt{2\alpha}}\ .
\end{equation}
Using these relations one can represent ${\cal V}(\phi)$ as follows:
\begin{eqnarray}
 {\cal V}(\phi) \sim - 3\beta B M_5^3   v \exp \left(-{\alpha m  \phi C^2\over \sqrt {6\beta \alpha  } M_p\, D }\right)  \ .
\label{propV2}
\end{eqnarray}
Here we will concentrate on the absolute value of the coefficient
${\alpha m  \phi C^2\over \sqrt {6\beta \alpha  } M_p\, D }$ in the
exponent. Using the inequalities   $C > D$, $C > \alpha M_5^{-1}$, and
$\alpha m M_5^{-1} \gtrsim 120$, one finds
\begin{eqnarray}
{\alpha m  \phi C^2\over \sqrt {6\beta \alpha  } M_p\, D } > {\alpha m
\phi C \over \sqrt {6\beta \alpha  } M_p  } > {50  \phi   \over  M_p
}\sqrt {\alpha \over\beta  } \ . \label{propV3}
\end{eqnarray}
Note that $\alpha \gg \beta$. Thus the factor in the exponent in the
expression for ${\cal V}(\phi)$, Eq. (\ref{propV2}), is much greater than
${50  \phi   \over  M_p  }$ for any choice of parameters in the
pyrotechnic scenario. In particular, if one takes $\alpha/\beta \sim 2500$
as in \cite{KOST}, one finds that the factor in the exponent must be
greater than ${2500  \phi   \over  M_p  }$.

\

\end{document}